\documentclass[preprint,12pt,authoryear]{elsarticle}

\usepackage{amssymb}
\usepackage{amsmath}

\usepackage[utf8]{inputenc}
\usepackage{csquotes}
\usepackage{macros}
\usepackage{url}
\usepackage{hyperref}
\usepackage{amsmath}
\usepackage{amssymb}
\usepackage{empheq} 
\usepackage{mdframed} 
\usepackage{amsfonts}
\usepackage{amssymb}
\usepackage{graphicx}
\usepackage{float}
\usepackage{lipsum}
\usepackage{xcolor}
\usepackage{appendix}
\newtheorem{remark}{Remark}
\newtheorem{assumption}{Assumption}
\usepackage{subcaption}
\usepackage{tikz}
\usepackage{tikz-3dplot}
\usetikzlibrary{arrows.meta, decorations.pathmorphing}
\journal{Computer Methods in Applied Mechanics and Engineering}

\begin{document}

\begin{frontmatter}



\title{Lie Group Variational Integrator for the Geometrically Exact Rod with Circular Cross-Section Incorporating Cross-Sectional Deformation} 


\author[SysCon]{Srishti Siddharth\corref{cor1}}\cortext[cor1]{Corresponding author}
\ead{214230005@iitb.ac.in}
\author[SysCon]{Vivek Natarajan\fnref{label2}}
\ead{vivek.natarajan@iitb.ac.in}
\author[SysCon]{ Ravi N. Banavar\fnref{label3}}
\ead{banavar@iitb.ac.in}
\affiliation[SysCon]{organization={Centre for Systems and Control, Indian Institute of Technology Bombay},
            addressline={Powai}, 
            city={Mumbai},
            postcode={400076}, 
            state={Maharashtra},
            country={India}}

\begin{abstract}
In this paper, we derive the continuous space-time equations of motion of a three-dimensional geometrically exact rod, or the Cosserat rod, incorporating planar cross-sectional deformation. 
We then adopt the Lie group variational integrator technique to obtain a discrete model of the rod incorporating both rotational motion and cross-sectional deformation as well. The resulting discrete model possesses several desirable features: it ensures volume conservation of the discrete elements by considering cross-sectional deformation through a local dilatation factor,  it demonstrates the beneficial properties associated with the variational integrator technique, such as the preservation of the rotational configuration, and energy conservation with a bounded error. 
An exhaustive set of numerical results under various initial conditions of the rod demonstrates the efficacy of the model in replicating the physics of the system. 
\end{abstract}




\end{frontmatter}



\section{Introduction}
\label{sec:intro}
In many engineering and scientific applications, geometrically exact rods are used to obtain accurate models for highly deformable slender rods. Such flexible structures find applications in continuum robots \cite{Gravagne,Cosimo_2023}, DNA strands experiencing various mechanical forces \cite{Forth_2008} and flexible microrobots for endovascular surgery \cite{Bailly_2011}. The geometrically exact rod \cite{Antman} or the Cosserat rod provides a mathematically elegant and high-fidelity model for such slender rods undergoing large deformations, unlike Euler-Bernoulli and Timoshenko beam theories, which account for small deformations. 

The variational formulation of the geometrically exact rod and its numerical aspects were presented in the extensive work of Simo et al. \cite{Simo_1985,Simo_1986,Simo_1988}. The essential features of the discretisation scheme proposed in these papers are the implicit algorithms to discretise the rotational dynamics along with the strain discretisation rules which preserve their properties after
discretisation. The book by Antman \cite{Antman} presents the theory of planar and three-dimensional geometrically exact rods. The modern treatment of geometrically exact rods is directed towards algorithm design and implementation based on the special Cosserat rod theory, which describes the motion of the rod using the position of its centerline and the orientation of the rigid cross-sections attached at each material point on the centerline. There are various Finite element method (FEM) approaches which are used to obtain such numerical models as well. The work in \cite{Boyer_2004} focuses on obtaining numerical models which account for finite bending and torsion. Other FEM based numerical integrators are provided in \cite{Cao_2006,Sonneville_2014,Zupan_2003}. 

The Cosserat rod approach has also been adopted by the graphics and mechanics community. In this setting, work by Bergou et al. \cite{Bergou} and Gazzola et al. \cite{gazzola_2018} proposed \enquote{discrete elastic rods}, which describe the motion of a discrete rod by the position of the nodes and orientation of the discrete edges connecting the nodes. The spatial discretisation employed combines finite difference approximations for various field quantities with discrete averaging of the rod's rotational strains. For temporal discretisation, the authors utilise a symplectic, second-order accurate Verlet integration scheme. Through the application of discrete differential geometry (DDG), the authors demonstrate various interesting physical phenomena, such as the formation of knots, plectonemes and solenoids. 

The soft robotics community has proposed various numerical models based on the Cosserat rod theory. One of the prevalent numerical approaches in this field is the shooting-based approach. The nonlinear partial differential equations describing the rod dynamics are semidiscretised in time using the Backward Difference Formula of order $\alpha$ (BDF-$\alpha$), followed by shooting method-based iterations to solve the semidiscretised ODEs with boundary conditions \cite{Sahrifi_Walker_2021,Till_2019}. Most of the numerical techniques described above are based on absolute parameterisation of the Cosserat rod. Recently, Boyer et al. \cite{Boyer_Renda_2021} have proposed a strain-based parametrisation to describe the motion of continuum robots which results in the statics and dynamics having a similar structure to that of rigid link robots.  

The nonlinear PDEs describing the dynamics of the rod evolve on the manifold $\R^3 \times \SOthree$, thus requiring numerical models which preserve the geometry of this deformation. 
Geometric numerical integration techniques preserve mechanical properties of a system such
as energy, momentum and configuration geometry, apart from being numerically efficient \cite{Hairer2004}. Variational integrators are a class of geometric numerical integrators which are based on the theory of discrete mechanics \cite{Marsden_West_2001}. The development of discrete mechanics and variational integrators presented by Marsden and West \cite{Marsden_West_2001} is widely used to obtain numerically efficient algorithms for a large class of mechanical systems. The structure-preserving algorithms have the advantage of providing numerically accurate and fast simulations with large time steps, along with stable long-time behaviour and energy conservation with a bounded error. The variational integrators for mechanical systems evolving on Lie groups are called Lie group variational integrators, abbreviated as LGVI. The LGVI technique has been extensively applied to the study of dynamics of various rigid body systems such as the 3-D pendulum \cite{Lee_3D_pendulum}, quadrotor \cite{Fan_2015}, spacecraft \cite{Lee_spacecraft}, etc. Recently, mechanical systems discretised by the variational integrator scheme have become a widespread option for obtaining discrete optimal control laws \cite{junge2005,Marsden_DGOC,Lee_2008} as the discrete mechanics and optimal control (DMOC) approach produces discrete optimal solutions which inherently possess characteristic properties of the mechanical system.   

A Lie group variational integrator for the three-dimensional exact beam dynamics was introduced by Demoures et al. (2015) \cite{Demoures2015}. The discrete rod dynamics thus obtained evolve on the manifold $\R^3 \times \SOthree $ at the discrete level and exhibit long-time conservation of the approximate energy as well as objectivity of the discrete strain measures. A multisymplectic Lie group variational integrator for the Cosserat rod is provided by Demoures et al. (2016) \cite{Demoures_Ratiu_2016} and Chen et al. (2022) \cite{Chen_2022}. A Lie group variational integrator which employs quaternions to describe the orientation of the rod is presented by Hante et al. \cite{Hante} and Tumiotto and Arnold \cite{Arnold_2024}. 

The theory and numerical models of the geometrically exact rod described above assume that the cross-sections remain rigid during deformation. There are a few studies present in the literature which explore the cross-sectional deformation in detail. In \cite{Kumar_Mukherjee_2010}, the authors have accounted for this effect by assuming that the directors do not remain orthonormal. In addition to considering the effect of Poisson's ratio to account for cross-sectional inflation due to stretching, they include additional strain measures to capture the cross-sectional deformation, thereby increasing the order of the ODE which describes the statics of the rod. In a similar development in \cite{sun2024}, the authors have included an inflation ratio (which results in an additional strain variable) in the deformation map and derive the equations of motion using the deformation gradient approach. In \cite{Tunay2013}, the author has considered quaternions to represent the orientation of the cross-sections. However, the quaternions are not restricted to be of unit norm, resulting in a rod model which accounts for inflation of the cross-section.

\textit{Contributions of the paper:} The contributions of this paper are twofold. The first contribution is the derivation of the three-dimensional equations of motion of the Cosserat rod incorporating planar cross-sectional deformation under the assumption that circular cross-sections of the cylindrical rod remain circular after deformation. This assumption is also made in Gazzola et al. \cite{gazzola_2018}, one of the articles we compare our results to. The authors of \cite{gazzola_2018} derive the equations of motion using a Newtonian approach, while we adopt a Lagrangian approach, incorporating the effects of cross-sectional deformation on the geometrical properties of the rod. 
The second contribution of the paper is the development and implementation of the discrete rod model using the Lie group variational integrator scheme. We build on the discrete model of the rod derived by Demoures et al. \cite{Demoures2015} by incorporating the new feature - cross-sectional deformation of the rod using a local dilatation factor. We present three numerical experiments to demonstrate the long-time approximate energy conservation and structure-preserving properties of the discrete model, while also ensuring the conservation of volume of the discrete elements of the rod. 

\textit{Outline of the paper: } In Section \ref{sec:notation_math_prelim}, we present 
a brief review of a few Lie group identities. In Section \ref{sec:CR_theory}, we derive the continuous space-time equations of motion of the three-dimensional Cosserat rod including the cross-sectional deformation and provide a comparison with the standard Cosserat rod theory presented in \cite{Demoures2015}. We then develop the discrete equations of motion of the proposed rod model by the Lie group variational integrator technique in Section \ref{sec:lgvi_modified_rod}. In Section \ref{sec:numerical_results}, we present and discuss three numerical experiments to validate the effects of including cross-sectional deformation on the dynamics of the rod. Section \ref{sec:conclusion_and_future_work} presents concluding remarks and future work.  
\section{Lie group preliminaries}
\label{sec:notation_math_prelim}
In this section, we recall definitions from the theory of Lie groups. \\
Let $\identitymat$ denote the identity element in $\R^{3 \times 3}$. The three-dimensional special orthogonal group $\SOthree$ is the set
$\{\rot \in \R^{3 \times 3} | \rot^T \rot = \rot \rot^T = \identitymat, \text{det} \rot = 1 \}$ with matrix multiplication as the group operation. 
The associated Lie algebra, $\sothree$, is the vector space of $3 \times 3$ real skew-symmetric matrices. For $\angvel \in \R^3$, we define $\widehat{\angvel} \in \sothree$ via the \enquote{hat} map as follows: 
\begin{equation}
        \R^{3} \ni {\angvel} = \begin{bmatrix}
            \angvel_1 \\
            \angvel_2 \\
            \angvel_3
        \end{bmatrix} \mapsto \widehat{\angvel} = \begin{bmatrix}
            0 & -{\angvel}_3 & {\angvel}_2 \\
            {\angvel}_3 & 0 & -{\angvel}_1 \\
            - {\angvel}_2 & {\angvel}_1 & 0
        \end{bmatrix} \in \sothree,
    \end{equation}
and the inverse of the hat map is the \enquote{vee} map $(\cdot)^{\vee}$ from $\sothree$ to $\R^3$ 
so that 
\begin{equation}
 (\widehat{\angvel})^{\vee} =  {\angvel}.
\end{equation}
The antisymmetric map $(\cdot)^{A}:\R^{3 \times 3} \to \sothree $ maps a matrix $M \in \R^{3 \times 3}$ to its skew-symmetric part, i.e. 
\begin{equation}
     (M)^{A} = \frac{M-M^T}{2}.
\end{equation}
The Cayley map $\textrm{cay}: \sothree \to \SOthree$ is defined as
\begin{equation}
      \textrm{cay} (\widehat{M}) =\Bigl(I_3 - \frac{\widehat{M}}{2}\Bigr)^{-1} \Bigl(I_3 + \frac{\widehat{M}}{2}\Bigr)
        \label{eq:cayley}
    \end{equation}
    for $\widehat{M} \in \sothree$ and its inverse is $\textrm{cay}^{-1}: \SOthree \to \sothree$ defined as
     \begin{equation}
       \invcay(\rot) = 2 (\rot-I_3)(\rot+I_3)^{-1} 
        \label{eq:inv_cayley}
    \end{equation}
for $R \in \SOthree$, see \cite{Iserles_2000} for details.   
\section{Description of the three-dimensional Cosserat rod incorporating cross-sectional deformation}
\label{sec:CR_theory}
\subsection{Kinematics of the rod}
\begin{figure}[H]
    \centering
    \includegraphics[width=0.8\linewidth]{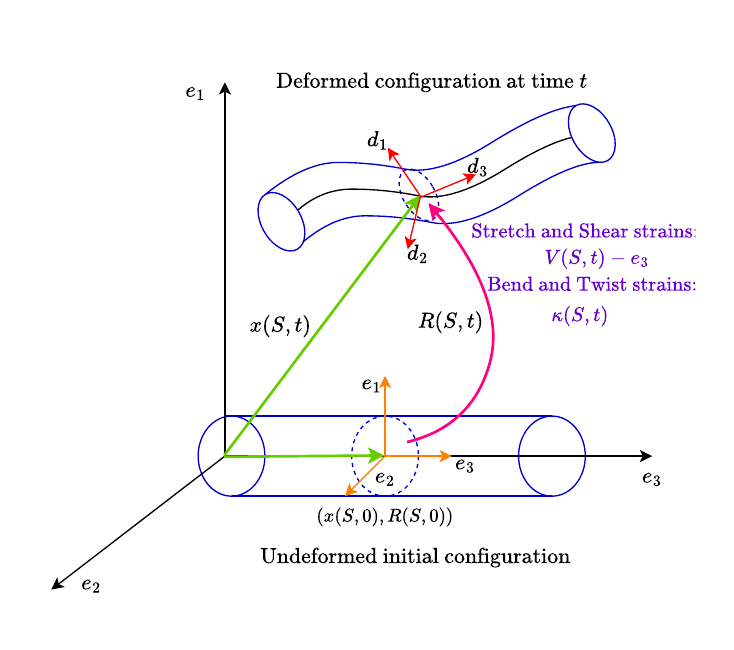}
    \caption{Deformation of the three-dimensional geometrically exact rod.}
    \label{fig:cosserat_rod}
\end{figure}

The geometrically exact rod or the Cosserat rod takes into account the following deformations: stretching, shearing, bending and twisting. The configuration of the rod is completely described at any given time instant by the position of the material points lying on the centerline and the orientation of the cross-sections attached to these points. In this paper, we take the initial reference configuration of the rod to be the one in which the rod is straight (zero curvature) and is undeformed. Let $L$ be the length of the undeformed rod and let $S \in [0, L]$ be the arc-length parameter which identifies the material points on the centerline of the rod. The position of the material point at $S$ at any time instant $t$,  in the inertial  orthonormal frame $\{{e}_1,{e}_2,{e}_3\}$, is denoted by $\x (S,t) \in \R^{3}$. Here ${e}_1,{e}_2,{e}_3$ are the standard unit vectors in $\R^3$. In the reference configuration, the rod is supposed to be located along the ${e}_3$ axis. In this work, we assume that the cross-section of the Cosserat rod remains circular at all time instants. Let $d_1^{S}$ and $d_2^{S}$ be two vectors parallel to $e_1$ and $e_2$, respectively, lying in the cross-section at the material point $S$ in the reference configuration. The orientation of this cross-section at a time instant $t$ is determined by the directions of $d_1^{S}$ and $d_2^{S}$ at that time instant, which are denoted using the unit vectors $d_1(S,t)$ and $d_2(S,t)$, respectively. We assume that $d_1(S,t)$ and $d_2(S,t)$ are orthogonal to each other at all time instants. The orthonormal moving frame (also known as body-fixed or material frame) at each $S$ is given by $\{{d}_1 (S,t),{d}_2 (S,t), {d}_3 (S,t) \}$.  Let $\rot (S,t) \in \SOthree$ represent the orientation of the cross-section at the material point $S$. Then we have the relation 
\begin{equation}
    {d}_i (S,t) = \rot (S,t) {e}_i, \quad i= 1,2,3.
\end{equation}
Figure \ref{fig:cosserat_rod} shows the Cosserat rod and the various frames used to describe the pose of the rod.\\
The configuration space of the rod is defined as
\begin{equation}
    Q = \mathcal{C}^{\infty} ([0,L], \R^{3} \times \SOthree)
\end{equation}
with appropriate boundary conditions.\hspace{0.3em}The configuration of the material point at $S$ at time instant $t$ is given by
\begin{equation}
    p (S,t) = (\x (S,t), \rot (S,t)). 
\end{equation}
In addition to the configuration variables described above, the kinematics of the rod is described by its linear and angular velocities. The linear velocity of the rod, denoted as ${\linvel} (S,t) \in \R^{3}$ is given by
\begin{equation}
    {\linvel} (S,t) = \frac{\partial \x (S,t)}{\partial t}. 
\end{equation}
In this paper,  $\Dot{(\cdot)}$ and $\frac{\partial}{\partial t}(\cdot)$ will be used to denote the time derivative. The angular velocity of the rod, denoted as $\angvel(S,t) \in \R^{3}$ is given by
\begin{equation}
    \angvel (S,t) = (\rot^T (S,t) \Dot{\rot} (S,t) )^{\vee}
\end{equation}
which can be rewritten using the \enquote{hat} notation as
\begin{equation}
    \Dot{\rot}(S,t) = \rot(S,t) \widehat{\angvel} (S,t).
\end{equation}
After considering the kinematics of the rod, the equations of motion can be derived either by applying Newton's second law of motion to an infinitesimal element or by finding the Euler-Lagrange equations associated with its Lagrangian. We will adopt the latter method since it provides the formulation of the equations of motion based on kinetic and strain potential energies of the rod, which will be useful to formulate the discrete dynamics using the Lie Group Variational Integrator \cite{Demoures2015}. 
\subsection{Planar cross-sectional deformation }
\label{sec:area_deformation}
\begin{figure}[H]
    \centering
    \includegraphics[width=0.8\linewidth]{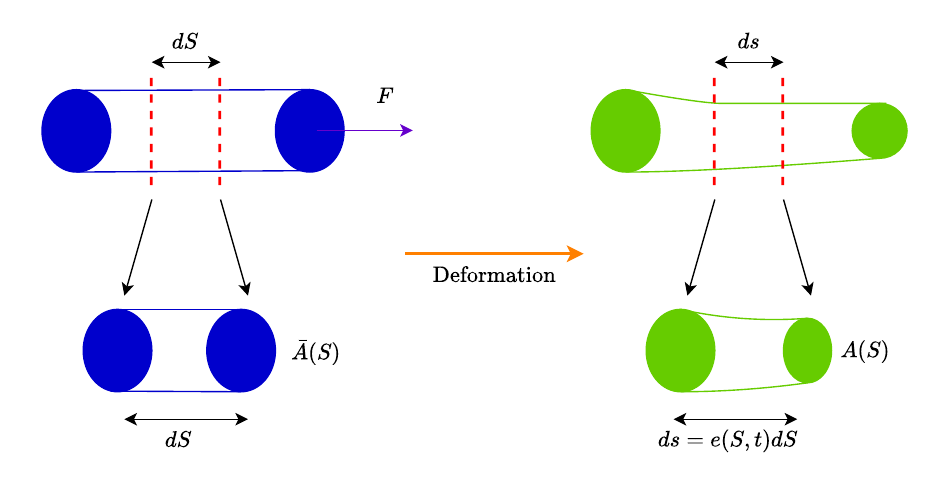}
    \caption{The cross-sectional deformation of the rod due to external force. }
    \label{fig:deformed_rod}
\end{figure}

In the special theory of Cosserat rod presented in Demoures et al. (2015) \cite{Demoures2015}, the cross-sections are assumed to be rigid for all kinds of deformations. This assumption permits limited deformations of the rod, although in many modern-day applications, deformations such as axial stretch, bending and twist change the cross-sectional area. 
In this section, we look at the planar deformation of circular cross-section of the rod due to extension, compression and bending. This will allow us to capture many interesting physical phenomena of the rod. In this section, we derive the equations of motion from a Lagrangian viewpoint, incorporating the cross-sectional deformation introduced in Gazzola et al. (2018) \cite{gazzola_2018}. 
\begin{assumption}
  The main assumption made to arrive at the new model is that the circular cross-section remains circular after deformation. This assumption will allow us to model deformation up to $30 \%$ of axial strain, as explained in \cite{gazzola_2018}. It should be noted that we have not introduced any additional strain variables to incorporate cross-sectional deformation, unlike other studies found in the literature.  
\end{assumption}
\begin{assumption}
    The material of the rod is assumed to be homogeneous, incompressible, isotropic, hyperelastic and is described with a linear stress-strain relationship.
    \label{assumption2}
\end{assumption}
Let $s \in \R$ be the current arc-length parameter due to the extension or compression of the rod. It is defined by the map
\begin{equation}
    S \mapsto s(S) = \int_{0}^{S} \norm{ \frac{\partial \x (u,t)}{\partial u} }  \,du,
\end{equation}
where $u$ is a dummy variable of integration \cite{Stewart}. Then, the local dilatation factor $e(S,t) \in \R$ is given as
\begin{equation}
    e(S,t) = \frac{ds}{dS} =  \norm{ \frac{\partial \x (S,t)}{\partial S} }.
    \label{eq:e_factor}
\end{equation}
Here, $0<e(S,t)<1$ and $e(S,t)>1$ represent compression and elongation of the infinitesimal rod element, $dS$, respectively, while $e(S,t) = 1, \forall S \in [0, L]$ represents an unstretched rod.  Since the mass of the infinitesimal element of the rod is conserved, therefore, we have 
\begin{equation}
    \rho \area dS = \rho A(S) ds \implies A(S) = \frac{\area}{e(S,t)}, 
    \label{eq:modified_area}
\end{equation}
where $\Bar{A} (S)$ and $A(S)$ are the cross-sectional area at $S$ in the initial and current configurations, respectively, as shown in Figure \ref{fig:deformed_rod} and $\rho$ is the density of the rod. \eqref{eq:modified_area} incorporates cross-sectional deformation while enforcing volume conservation of the infinitesimal element of the rod. 
\subsection{Lagrangian of the rod}
\label{sec:lagrangian}
In this section, we will describe the Lagrangian of the Cosserat rod which accounts for cross-sectional deformation. 

The Lagrangian of the rod, $\mathcal{L}: TQ \to \R$, consists of the kinetic and potential energies which are described as follows:
\begin{enumerate}
    \item Kinetic energy: The total kinetic energy of the rod is the sum of translational energy, denoted by $\mathcal{T}_l$ and rotational energy, denoted by $\mathcal{T}_r$.
    \begin{enumerate}
        \item  Translational kinetic energy:\\ The translational kinetic energy of an infinitesimal element $dS$ of the rod of mass $dm(t)$ at time $t$ is given as
    \begin{align}
    \begin{split}
         d\mathcal{T}_l(t) = dm(t)\Dot{\x}^T(S,t)\Dot{\x}(S,t) & = dm(0) \Dot{\x}^T(S,t)\Dot{\x}(S,t) \\
         &= \rho \Bar{A}(S) \Dot{\x}^T(S,t)\Dot{\x}(S,t) dS 
    \end{split}
    \end{align}
    due to the conservation of mass given in (\ref{eq:modified_area}). Thus, the translational kinetic energy is independent of $e(S,t)$ and continues to be the same as for the standard Cosserat rod \cite{Demoures2015}, which is given as
    \begin{equation}
        \mathcal{T}_l (t) = \int_{0}^{L} \frac{1}{2} \rho \Bar{A}(S) \Dot{\x}^T(S,t) \Dot{\x} (S,t) \,dS.
        \label{eq:Linear_KE}
    \end{equation}
   \item Rotational kinetic energy:\\
   The rotational kinetic energy of the element $dS$ of the rod is given as
    \begin{equation}
        d\mathcal{T}_r (t)= \frac{1}{2} \angvel^T(S,t) {J}(S,t) \angvel (S,t),
    \end{equation}
where ${J}(S,t)$ is the mass moment of inertia per unit length of the rod at $S$. 

For a slender rod, the mass moment of inertia is expressed in terms of the second moment of inertia. The second moment of inertia of the rod is given by $\secondMoI (S) = \text{diag} (\mathcal{I}_1 (S), \mathcal{I}_2 (S),  \mathcal{I}_3 (S))$. For a circular cross-section, we have
        \begin{equation}
            \mathcal{I}_1 (S) = \mathcal{I}_2 (S) = \frac{A^2 (S)}{4 \pi},  \mathcal{I}_3 (S) = \frac{A^2 (S)}{2 \pi}. 
        \end{equation} 
Due to the effect of the dilatation factor on the cross-sectional area in (\ref{eq:modified_area}), the second moment of inertia is given by 
\begin{equation}
    \secondMoI (S,t) = \frac{\overline{\secondMoI} (S)}{e^2(S,t)}.
    \label{eq:modified_second_MoI}
\end{equation}
Note that $\overline{(.)}$ represents various physical quantities in the reference configuration.

Thus, the mass moment of inertia of the deformed element at $S$ can be written as
   \begin{equation}
        {J} (S,t) = \rho \frac{\overline{\secondMoI}(S)}{e^2(S,t)} e(S,t) dS = \rho \frac{\overline{\secondMoI}(S)}{e(S,t)} dS
    \end{equation}
    and the rotational kinetic energy is given by
\begin{equation}
        \mathcal{T}_r (t) = \int_{0}^{L} \frac{1}{2} \rho {\angvel^T (S,t)} \frac{\overline{\secondMoI} (S)}{e(S,t)} \angvel(S,t) \,dS.
        \label{eq:modified_rot_ke}
\end{equation}
    \end{enumerate}
    \item Strain potential energy: 
    \begin{enumerate}
        \item Linear strain potential energy:
        The stretching and shearing strains of the rod are defined as
        \begin{equation}
            {V}(S,t) - {e}_3, \quad \text{where} \quad {V}(S,t) := \rot^T (S,t) \frac{\partial \x (S,t)}{\partial S} 
            \label{eq:stretch_strain}
        \end{equation}
        and 
        \begin{equation}
            {V}(S,t) - {e}_3 = \begin{bmatrix}
                \epsilon_{s_1} (S,t) \\
                \epsilon_{s_2} (S,t) \\
                \epsilon_{a} (S,t)
            \end{bmatrix}=\begin{bmatrix}
            v_1(S,t)\\v_2(S,t)\\v_3(S,t)-1
        \end{bmatrix}.
        \end{equation} $\epsilon_{s_1} = v_1(S,t) \in \R$ and $\epsilon_{s_2} = v_2(S,t) \in \R$ represent shearing strains along the $e_1$ and $e_2$ axes and $\epsilon_{a} = v_3(S,t)-1 \in \R$ represents the axial strain along the $e_3$ axis. In this paper, $\partialwrtS (\cdot)$ denotes the spatial derivative.

        Assuming a linear constitutive relationship, the linear stresses are given as
        \begin{equation}
            \sigma_{s_1} (S,t) = G v_1(S,t),
        \end{equation}
        \begin{equation}
            \sigma_{s_2} (S,t) = G v_2(S,t),
        \end{equation}
        \begin{equation}
            \sigma_{a} (S,t) = E (v_3(S,t) -1),
        \end{equation}
        where $\sigma_{s_1},\sigma_{s_2},\sigma_{a} \in \R$ are the shear and axial stresses, $E$ is the Young's modulus and $G$ is the shear modulus of the rod. The relation between the two moduli is given as
        \begin{equation}
            G = \frac{E}{2(1+\nu)},
        \end{equation}
        where $\nu$ is the Poisson's ratio of the rod. The  stretching and shearing stiffness of the rod is $\frac{\Bar{\stretchstiff}(S)}{e(S,t)}$, where $\Bar{\stretchstiff} (S) = \text{diag} (G \Bar{A}(S),G \Bar{A}(S)$ $, E \Bar{A}(S))$. This is due to the cross-sectional area deformation defined in (\ref{eq:modified_area}).
         \begin{assumption}
             We assume that the material-specific parameters $E, G$ and $\nu$ are not affected by $e(S,t)$ and are assumed to be constants of the rod as it undergoes deformation.  
         \end{assumption}

        Accounting for the cross-sectional deformation in (\ref{eq:modified_area}), the forces generated due to the shear and axial stresses are given by
        \begin{equation}
        \begin{aligned}
             F_{s_1} (S,t) &= f_1(\epsilon_{s_1} (S,t)) =  \sigma_{s_1} (S,t) A (S) \\
             & =   \frac{G \area}{e(S,t)} \epsilon_{s_1} (S,t)=  \frac{G \area}{e(S,t)} v_1(S,t),
        \end{aligned}
        \label{eq:fs1}
        \end{equation}
        \begin{equation}
        \begin{aligned}
             F_{s_2} (S,t) &= f_2(\epsilon_{s_2}(S,t)) = \sigma_{s_2} (S,t) A (S) \\
            & = \frac{G \area}{e(S,t)} \epsilon_{s_2}(S,t)  =\frac{G \area}{e(S,t)} v_2(S,t),
        \end{aligned}
        \label{eq:fs2}
        \end{equation}
        \begin{equation}
        \begin{aligned}
             F_{a} (S,t) &= f_3(\epsilon_a (S,t)) = \sigma_{a} (S,t) A (S) \\
             &= \frac{E \area}{e(S,t)} \epsilon_a (S,t) =  \frac{E \area}{e(S,t)} (v_3(S,t) -1),
        \end{aligned}
        \label{eq:fa}
        \end{equation}
        where $F_{s_1},F_{s_2},F_{a}$ are the shear and axial forces.

        The work done on an infinitesimal element of the rod $dS$ to generate the linear strains is given as
        \begin{equation}
            \begin{aligned}
                dW_{\text{lin}} &=  dW_{s_1} +  dW_{s_2} + dW_{a} \\
                & = \bigl[ \int_{0}^{\epsilon_{s_1}} f_1 (\xi_{s_1})\,d\xi_{s_1} + \int_{0}^{\epsilon_{s_2}} f_2 (\xi_{s_2})\,d\xi_{s_2}  \\
                &+ \int_{0}^{\epsilon_{a}}f_3 (\xi_{a})\,d\xi_{a}  \bigr] \,dS,
            \end{aligned}
        \end{equation}
        where $\xi_{s_1},\xi_{s_2}, \xi_{a}$ are the dummy variables for shearing and stretching strains.
        The total work done by the forces $F_{s_1},F_{s_2}, F_{a}$, denoted by $W_{lin}$ is stored as the linear strain potential energy $\mathcal{U}_{lin}$. Thus,  
        \begin{equation}
        \begin{aligned}
             W_{lin} &= \mathcal{U}_{lin} = \int_{0}^{L} dW_{lin} \\
             &= \int_{0}^{L} \bigl[ \int_{0}^{\epsilon_{s_1}} f_1 (\xi_{s_1})\,d\xi_{s_1} + \int_{0}^{\epsilon_{s_2}} f_2 (\xi_{s_2})\,d\xi_{s_2} \\
             &+ \int_{0}^{\epsilon_{a}}f_3 (\xi_{a})\,d\xi_{a}  \bigr] dS.
        \end{aligned}
        \label{eq:PE_lin}
        \end{equation}
        \item Rotational strain potential energy: The bending and twisting strains of the rod, denoted by $\curvature (s,t) \in \R^{3}$ are defined as
    \begin{align}
    \begin{split}
         \curvature (s, t) &= \Bigl((\rot (s,t))^T \frac{\partial \rot (s,t)}{\partial s} \Bigr)^{\vee} \\
         &= \frac{1}{e(S,t)}\Bigl((\rot (S,t))^T \frac{\partial \rot (S,t)}{\partial S}\Bigr)^{\vee} \\
         &= \frac{\curvature (S, t)}{e(S,t)},
    \end{split}
\label{eq:modified_curvature_vector}
    \end{align}
where $ds = e(S,t)dS$ from (\ref{eq:e_factor}) and the angular strain $\curvature (S,t) = \begin{bmatrix}
    \curvature_1 (S,t) & \curvature_2 (S,t) & \curvature_3 (S,t)
\end{bmatrix}^T$ $\in \R^3$ is expressed in terms of the original arc-length parameter $S$.
Accounting for the cross-sectional deformation in (\ref{eq:modified_area}), the moments generated due to bending and twisting stresses are given as
\begin{equation}
    \begin{aligned}
        \tau_1 (S,t)  = f_4(\curvature_1 (S,t)) = \frac{E\Bar{\mathcal{I}}_1 (S)}{e^3(S,t)} \curvature_1 (S,t), 
    \end{aligned}
    \label{eq:fxi4}
\end{equation}
\begin{equation}
    \begin{aligned}
        \tau_2 (S,t)  = f_5(\curvature_2 (S,t)) = \frac{E\Bar{\mathcal{I}}_2 (S)}{e^3(S,t)} \curvature_2 (S,t), 
    \end{aligned}
    \label{eq:fxi5}
\end{equation}
\begin{equation}
    \begin{aligned}
        \tau_3 (S,t)  = f_6(\curvature_3 (S,t)) = \frac{G\Bar{\mathcal{I}}_3 (S)}{e^3(S,t)} \curvature_3 (S,t), 
    \end{aligned}
    \label{eq:fxi6}
\end{equation}
where $\tau_1,\tau_2$ are the stresses generated due to the bending strains about $e_1$ and $e_2$ axes, and $\tau_3$ is the stress generated due to twisting about the $e_3$ axis.
The angular stiffness of the rod is $\frac{\Bar{\bendstiff}(S)}{e^2(S,t)}$ where $\Bar{\bendstiff} (S) = \text{diag} (E\Bar{\mathcal{I}}_1 (S), E\Bar{\mathcal{I}}_2 (S), G\Bar{\mathcal{I}}_3 (S))$. This is due to the effect of cross-sectional deformation on the second moment of inertia given in (\ref{eq:modified_second_MoI}). 

The work done on an infinitesimal element of the rod $dS$ to generate the angular strains is given as
        \begin{equation}
            \begin{aligned}
                dW_{\text{ang}} &=  dW_{1} +  dW_{2} + dW_{3} \\
                & = \bigl[ \int_{0}^{\curvature_1} f_4 (\xi_{4})\,d\xi_{4} + \int_{0}^{\curvature_2} f_5 (\xi_{5})\,d\xi_{5} \\
                &+ \int_{0}^{\curvature_3} f_6 (\xi_{6})\,d\xi_{6}  \bigr] \,dS,
            \end{aligned}
        \end{equation}
        where $\xi_{4},\xi_{5},\xi_{6}$ are the dummy variables for bending and twisting strains.

        The total work done by the moments $\tau_1,\tau_2,\tau_3$, denoted by $W_{ang}$ is stored as the rotational strain potential energy $\mathcal{U}_{ang}$. Thus,  
        \begin{equation}
        \begin{aligned}
             W_{ang} &= \mathcal{U}_{ang} = \int_{0}^{L} dW_{ang} \\
             &= \int_{0}^{L} \bigl[ \int_{0}^{\curvature_1} f_4 (\xi_{4})\,d\xi_{4} + \int_{0}^{\curvature_2} f_5 (\xi_{5})\,d\xi_{5} \\
             &+ \int_{0}^{\curvature_3} f_6 (\xi_{6})\,d\xi_{6}  \bigr] \, dS.
        \end{aligned}
        \label{eq:PE_ang}
        \end{equation}
    \end{enumerate}
The total strain potential energy of the rod, obtained by adding (\ref{eq:PE_lin}) and (\ref{eq:PE_ang}) is given as
\begin{equation}
    \mathcal{U} (t) = \mathcal{U}_{lin} + \mathcal{U}_{ang}.
    \label{eq:PE}
\end{equation}
\begin{remark}
    The linear and rotational strain potential energies given in (\ref{eq:PE_lin}) and (\ref{eq:PE_ang}) cannot be integrated analytically as the stiffness matrices depend on the dilatation factor. Equations (\ref{eq:PE_lin}) and (\ref{eq:PE_ang}) reduce to the standard quadratic form
    \begin{equation}
        \int_{0}^{L} \frac{1}{2} ({V}(S,t) - {e}_3)^T {\Bar{\stretchstiff}(S)} ({V}(S,t) - {e}_3) +  \frac{1}{2} \curvature^T (S,t) {\Bar{\bendstiff}(S)} \curvature (S,t) \,dS  
    \end{equation}
    if the cross-sectional deformation is not considered.
\end{remark}

\subsection{The governing equations of motion}
\label{eq:equations_of_motion}
The equations of motion of the rod for $t \in [0,T]$ are derived by computing the variation of the action integral given by
\begin{equation}
    \mathcal{S} = \int_{0}^{T} \mathcal{T}_l (t) + \mathcal{T}_r (t) - \mathcal{U} (t) \, dt,
    \label{eq:action_int}
\end{equation}
where $\mathcal{T}_l (t)$, $\mathcal{T}_r (t)$ and $\mathcal{U} (t)$ are given in (\ref{eq:Linear_KE}), (\ref{eq:modified_rot_ke}) and (\ref{eq:PE}), respectively.

The equations of motion of the rod incorporating the cross-sectional deformation in the presence of non-conservative, distributed external force $\mathcal{F} (S,t) \in \R^3$ and moment $\tau (S,t) \in \R^3$ are derived by applying the Lagrange- D'Almebert principle to the action integral given above. The detailed proof of the continuous space-time equations of motion of the Cosserat rod including cross-sectional deformation has been provided in \ref{app:derivation_CT}.
Thus, the translational equation of motion is given as
\begin{empheq}[box=\fbox]{equation}
    \begin{aligned}
        \rho \Bar{A}(S)\Ddot{\x} (S,t) &=  \partialwrtS \Bigl(\rot(S,t) \frac{\Bar{\stretchstiff}(S)}{e(S,t)}\axialstrain \Bigr)  + \mathcal{F}(S,t) 
    \end{aligned}
     \label{eq:lin_eom}
\end{empheq}
%
%
and the rotational equation of motion is given as 
\begin{empheq}[box=\fbox]{gather}
\begin{aligned}
\rho \frac{\overline{\secondMoI}(S)}{e(S,t)} \Dot{\angvel}(S,t)
&
= \partialwrtS \Bigl(
\frac{\Bar{\bendstiff}(S)\curvature(S,t)}{e^3(S,t)}
\Bigr) \\
&+ \frac{\curvature(S,t)\times
\bigl(\Bar{\bendstiff}(S)\curvature(S,t)\bigr)}{e^3(S,t)} \\
&\quad + \Bigl(\rot^T(S,t) \frac{\partial \x(S,t)}{\partial S}\Bigr)\times
\frac{\Bar{\stretchstiff}(S)}{e(S,t)}\axialstrain\\
&\quad + \rho \frac{\overline{\secondMoI}(S)}{e(S,t)}
\angvel(S,t)\times\angvel(S,t) \\
&\quad + \rho \frac{\overline{\secondMoI}(S)}{e^2(S,t)}
\angvel(S,t)\Dot{e}(S,t) + \tau(S,t)
\end{aligned}
\label{eq:rot_eom}
\end{empheq}
$\forall S \in [0,L]$.
\begin{remark}
The solution of the set of PDEs (\ref{eq:lin_eom}) and (\ref{eq:rot_eom}) requires initial configuration of the rod along with boundary conditions. For example, the boundary conditions for a cantilever rod with an  external force, $f(L,t) \in \R^3$ and a moment, $\tau(L,t) \in \R^{3} $ acting on the tip are 
\begin{equation}
    \x(0,t) = 0,\quad \rot(0,t) = \identitymat, \quad \forall t \geq 0, 
    \label{eq:fixed_bc}
\end{equation}
\begin{align}
        \begin{split}
            &\left. \rot (S,t)\frac{\Bar{\stretchstiff}(S)}{e(S,t)} \axialstrain 
 \right\vert_{S=L} = f(L,t),\\
    \end{split}
      \label{eq:lin_bc}
\end{align}
\begin{equation}
    \left. \frac{\Bar{\bendstiff}(S)}{e^3 (S,t)}\curvature (S,t)\right\vert_{S=L} = \tau(L,t)
    \label{eq:rot_bc}
\end{equation}
$\forall t \geq 0$.
\end{remark}
\begin{remark}
   The translational equation of motion provided in \cite{Demoures2015}, which does not account for cross-sectional deformation is given as 
%
\begin{equation}
    \rho \Bar{A}(S) \Ddot{\x} (S,t) - \rot (S,t) n'(S,t) - \rot (S,t) (\curvature (S,t) \times n(S,t))  = \mathcal{F} (S,t) 
    \label{eq:eom_lin_stress}
\end{equation}
and the rotational equation in \cite{Demoures2015} is given as
\begin{align}
\begin{split}
    \Bar{J} (S) \Dot{\angvel} (S,t) + \angvel (S,t) \times \Bar{J}(S) \angvel (S,t) + n (S,t)  \times {V} (S,t) \\
    -\curvature (S,t) \times m (S,t) -m'(S,t) = \tau (S,t),   \end{split}
    \label{eq:eom_rot_stress}
\end{align}
where 
\begin{equation}
    n (S,t):= 
    \frac{\partial \mathcal{U}}{\partial V} =  \Bar{\stretchstiff} \axialstrain \in \R^3 
\end{equation}
and 
\begin{equation}
    m (S,t):=  \frac{\partial \mathcal{U}}{\partial \curvature} =  \Bar{\bendstiff} \curvature (S,t) \in \R^3.
\end{equation} 
%
%
The components of $n(S,t)$ along $e_1$ and $e_2$ are the shear stresses, and the component along $e_3$ is the axial stress of the rod. Similarly, the components of $m (S,t)$ along $e_1$ and $e_2$ are the bending momenta, and the component along $e_3$ is the twisting moment of the rod. 

The boundary conditions for the cantilever rod without the additional cross-sectional deformation are given as
\begin{align}
\begin{split}
     & \x (0,t) = 0  \\
     & \rot (0,t) = \identitymat , \quad \forall t \in [0,T]
\end{split}
\end{align}
and
\begin{align}
    \begin{split}
       & n(L,t) = (\rot(L,t))^T f(L,t) \\
       & m(L,t) = \tau(L,t) , \quad \forall t \in [0,T].
       \label{eq:BC_standard}
    \end{split}
\end{align} 
\end{remark}
Therefore, as compared to the standard Cosserat rod described in (\ref{eq:eom_lin_stress}) - (\ref{eq:BC_standard}), the model proposed in (\ref{eq:lin_eom}) - (\ref{eq:rot_bc}) is a consequence of the scaling of the geometrical and material features of the rod by the dilatation factor. 
\begin{remark}
  It is observed that the equations of motion (\ref{eq:lin_eom}) - (\ref{eq:rot_bc}), derived using a Lagrangian approach, are identical to those in \cite{gazzola_2018}, which have been derived using Newtonian techniques. 
\end{remark}

\end{enumerate}

\section{Lie group variational integrator for the rod incorporating cross-sectional deformation}
\label{sec:lgvi_modified_rod}
In this section, we proceed to derive the discrete model of the Cosserat rod incorporating cross-sectional deformation as explained in Section \ref{sec:CR_theory}. We will refer to the Cosserat rod described in \cite{Demoures2015} as the standard rod model and the Cosserat rod described in Section \ref{sec:CR_theory} as the modified rod model. In \cite{Marsden_West_2001}, the fundamental ideas of the variational integrator technique for a general configuration manifold have been presented. In this section, we will look at the various aspects of the Lie group variational integrator (LGVI) model of the 3-D Cosserat rod described in Section \ref{sec:CR_theory}. The discrete model for the standard Cosserat rod based on the LGVI technique was introduced in \cite{Demoures2015}. Our objective is to derive a discrete rod model which also accounts for cross-sectional deformation. Thus, our work combines the modified rod model and a structure-preserving discretisation scheme.  Based on the kinetic and strain potential energies, the LGVI model of the modified rod can be obtained in three steps: (1) formulating the spatially discretised Lagrangian of the rod, (2) finding the approximate linear and angular velocities of the rod by selecting two nearby configurations in the configuration space and obtaining the discrete-time approximation of the spatially discretised Lagrangian computed in (1), thus deriving the completely discretised Lagrangian, and (3) computing the variation of the discrete action sum. The discretisation approach is explained in the following sections.
\subsection{Spatial discretisation}
\label{sec:spatial_discretisation}
Let us consider that the rod has been spatially discretised into $\nodes$ intervals with a uniform step size denoted by $l_{\nodes}$. Thus, for a rod of undeformed length $L$, we have
    \begin{equation}
        l_{\nodes} = \frac{L}{\nodes}.
    \end{equation}
    Let us denote each spatial node with $\nodept \in [0,\nodes]$ such that $\nodept=0$ and $\nodept=\nodes $ denote the base and the tip of the rod, respectively. We will denote the position and orientation of the cross-section at each spatial node by $\x_\nodept$ and $\rot_\nodept$, respectively.\\
    Let us consider the interpolation of the configuration variables in the interval $\interval = (\nodept,\nodept+1)$, given in   \cite{Demoures2015}, as
    \begin{align}
        \begin{split}
            \rot(s)|_{\interval} := \rot_\nodept \exp{\Bigl(\frac{S}{l_\nodes} \widehat{\psi}_\nodept   \Bigr)}, \quad \x(S)|_{\interval} := \x_\nodept + \frac{S}{l_\nodes} \Delta \x_\nodept, 
        \end{split}
    \end{align}
    where $S \in [0,l_\nodes]$, $\Delta \x_\nodept = \x_{\nodept+1}-\x_\nodept$ and $\exp(\widehat{\psi}_\nodept) = {\rot}^{T}_\nodept \rot_{\nodept+1}$. These interpolations were introduced in \cite{Crisfield1999} to maintain the strain objectivity of the discrete strains.\\
    \textit{Translational kinetic energy}\\
    Recall from Section \ref{sec:lagrangian} that the translational kinetic energy of the rod at time $t$ is given as
        
        \begin{equation}
            \mathcal{T}_l = \int_{0}^{L} \frac{1}{2} \rho \area \Dot{\x}^T(S,t) \Dot{\x} (S,t) \, dS = \sum_{\nodept=0}^{\nodept= \nodes} \int_{\nodept}^{\nodept+1} \frac{1}{2} \rho \area \Dot{\x}^T(S,t) \Dot{\x} (S,t) \, dS.
        \end{equation}
        By approximating the integral on the right-hand side of the above equation by the trapezoidal rule, we get
        \begin{equation}
              \int_{\nodept}^{\nodept+1} \frac{1}{2} \rho \area \Dot{\x}^T(S,t) \Dot{\x} (S,t) \, dS \approx \frac{l_{\nodes}}{4} \rho \Bar{A} [||\Dot{\x}_{\nodept}||^{2}+||\Dot{\x}_{\nodept+1}||^{2}] = \mathcal{T}_{ld},
            \label{eq:lin_ke_sd}
        \end{equation}
        where $\mathcal{T}_{ld}$ is the discrete translational kinetic energy.\\
    \textit{Rotational kinetic energy}\\
    The rotational kinetic energy given in (\ref{eq:modified_rot_ke}), can be approximated in the spatial interval $[\nodept,\nodept+1]$ as
        \begin{equation}
              \int_{\nodept}^{\nodept+1} \frac{1}{2} \angvel^T (S,t) \frac{\Bar{\massMoI}}{e(S,t)} \angvel (S,t) \, dS \approx  \frac{l_{\nodes}}{4} \left[ \angvel^T_\nodept \frac{\Bar{\massMoI}_\nodept}{e_\nodept}\angvel_\nodept + \angvel^T_{\nodept+1 }\frac{\Bar{\massMoI}_{\nodept+1} }{e_{\nodept+1}}\angvel_{\nodept+1} \right] = \mathcal{T}_{rd},
              \label{eq:rot_ke_sd}
        \end{equation}
        where $\mathcal{T}_{rd}$ is the discrete rotational kinetic energy,  $\Bar{\massMoI}_\nodept  = \rho \Bar{\secondMoI}_\nodept$ is the mass moment of inertia per unit length of the rod, and $e_\nodept \in \R$  defined as
        \begin{equation}
            e_\nodept = \frac{||\x_{\nodept+1}-\x_\nodept||}{l_{\nodes}}  \quad \forall \quad \nodept = 0,1,...,\nodes-1,
            \label{eq:discrete_dilatation}
         \end{equation}
     is the discrete local dilatation factor obtained by spatially discretising (\ref{eq:e_factor}).\\
     \textit{Potential energy stored due to shearing and stretching strains}\\
     The shearing and stretching strains in the spatial interval $[\nodept,\nodept+1]$ can be approximated as
        \begin{equation}
            {V}_\nodept -{e_3} = \rot^T_\nodept \frac{\Delta \x_\nodept}{l_{\nodes}} - {e_3}, 
        \end{equation}
        where $\Delta \x_\nodept = \x_{\nodept+1}-\x_\nodept$. The linear strain  potential energy, given in (\ref{eq:PE_lin}) , can be approximated by the trapezoidal rule as
        \begin{equation}
            \begin{aligned}
                \mathcal{U}_{sd} &=\int_{\nodept}^{\nodept+1} \bigl[ \int_{0}^{\epsilon_{s_{1}}} f_1 (\xi_{s_1})\,d\xi_{s_1} + \int_{0}^{\epsilon_{s_{2}}} f_2 (\xi_{s_2})\,d\xi_{s_2} + \int_{0}^{\epsilon_{a}}f_3 (\xi_{a})\,d\xi_{a}  \bigr] \,dS \\
                & \approx \frac{l_{\nodes}}{2} \big[ \int_{0}^{\epsilon_{s_{1,\nodept}}} f_1 (\xi_{s_1})\,d\xi_{s_1} + \int_{0}^{\epsilon_{s_{2,\nodept}}} f_2 (\xi_{s_2})\,d\xi_{s_2} + \int_{0}^{\epsilon_{a,\nodept}}f_3 (\xi_{a})\,d\xi_{a} \\
                & \int_{0}^{\epsilon_{s_{1,\nodept+1}}} f_1 (\xi_{s_1})\,d\xi_{s_1} + \int_{0}^{\epsilon_{s_{2,\nodept+1}}} f_2 (\xi_{s_2})\,d\xi_{s_2} + \int_{0}^{\epsilon_{a,\nodept+1}}f_3 (\xi_{a})\,d\xi_{a}\bigr],
            \end{aligned}
            \label{eq:Usd}
        \end{equation}
where $\epsilon_{s_{1,\nodept}},\epsilon_{s_{2,\nodept}},\epsilon_{s_{a,\nodept}}$ are the shearing and stretching discrete strains at node $\nodept$ respectively.\\
\textit{Potential energy stored due to bending and twisting strains}\\
The bending and twisting strains in the spatial interval $[\nodept,\nodept+1]$ can be approximated as
        \begin{equation}
            \curvature_{\nodept} = \frac{\discpsi{\psi}{\nodept}}{l_{\nodes}} \in \R^3,
        \end{equation}
        where $\widehat{{\psi}}_{\nodept} = \exp^{-1}{(\rot^T_{\nodept} \rot_{\nodept+1})}$ is computed using the inverse Cayley transform given in (\ref{eq:inv_cayley}). Thus, the rotational potential energy, given in (\ref{eq:PE_ang}) can be approximated by the rectangular rule as 
        \begin{equation}
            \begin{aligned}
                \mathcal{U}_{bd} &= \int_{\nodept}^{\nodept+1} \Bigl[ \int_{0}^{\curvature_1} f_4 (\xi_{4})\,d\xi_{4} + \int_{0}^{\curvature_2} f_5 (\xi_{5})\,d\xi_{5} + \int_{0}^{\curvature_3} f_6 (\xi_{6})\,d\xi_{6}  \Bigr] \,dS \\
                & \approx {l_{\nodes}} \Bigl[ \int_{0}^{\curvature_{1,q}} f_4 (\xi_{4})\,d\xi_{4} + \int_{0}^{\curvature_{2,q}} f_5 (\xi_{5})\,d\xi_{5} + \int_{0}^{\curvature_{3,q}} f_6 (\xi_{6})\,d\xi_{6}  \Bigr],
            \end{aligned}
        \label{eq:Ubd}\end{equation}
 where $\curvature_{{1,\nodept}},\curvature_{{2,\nodept}},\curvature_{{a,\nodept}}$ are the bending and twisting discrete strains at node $\nodept$,  respectively.\\       
Therefore, the spatially discrete potential energy is given as
        \begin{equation}
            \mathcal{U}_{d} = \mathcal{U}_{sd} + \mathcal{U}_{bd}.
            \label{eq:pe_sd}
        \end{equation}
On combining (\ref{eq:lin_ke_sd}), (\ref{eq:rot_ke_sd}) and (\ref{eq:pe_sd}), the spatially discrete Lagrangian is defined as
    \begin{equation}
        L_{sd} := \sum_{\nodept = 0}^{\nodes} \mathcal{T}_{ld} + \mathcal{T}_{rd} - \mathcal{U}_{d}.
    \end{equation}
\subsection{Temporal discretisation}
Let us consider that the time interval $[0,T]$ is divided into $N$ intervals of uniform size $h$. Thus, we have
    \begin{equation}
        h = \frac{T}{N}.
    \end{equation}
    We will denote each time instant with $k \in [0,N]$ such that $k = 0$ and $k = N$ denote the initial and final time, respectively. We will denote the position and orientation of the cross-section at each spatial node $\nodept$ and time instant $k$ by $x^k_\nodept$ and $R^k_\nodept$,  respectively.\\
    \textit{Discrete linear velocity of the rod }\\
    The linear velocity is approximated by considering two nearby rod configurations in time. Thus, the discrete linear velocity is given as
    \begin{equation}
        \linvel(S,t) = \Dot{\x}(S,t) \approx \frac{\x^{k+1} (S)-\x^k (S)}{h}.
    \end{equation}
    On combining the discrete-space and discrete-time quantities, the completely discretised linear velocity is given as
    \begin{equation}
      \linvel_\nodept^k =  \Delta {\x}^k_\nodept = \frac{\x^{k+1}_\nodept-\x^k_\nodept}{h} \in \R^3
    \end{equation}
    for all $\nodept = 0,1,...,\nodes$.\\
 \textit{Discrete angular velocity of the rod}\\
 The discrete angular velocity has an interesting expression which arises due to the approximation of $\Dot{\rot}$. The angular velocity at node $\nodept$ can be approximated as
    \begin{equation}
        \widehat{\angvel}_\nodept = {\rot^T_\nodept}{\Dot{\rot}_\nodept} \approx {(\rot^k_\nodept)^T} \frac{\rot^{k+1}_\nodept-\rot^k_\nodept}{h} = \frac{{F}^k_\nodept-I_3}{h},
    \end{equation}
    where $F^k_\nodept \in \SOthree$ is the rotation update defined as
    \begin{equation}
           {F}^k_\nodept = (\rot^{k}_\nodept)^{T} {\rot^{k+1}_\nodept}.
    \end{equation}
    Thus, the angular velocity of the cross-section at spatial node $\nodept$ and time instant $k$ is given by
    \begin{equation}
        \widehat{\angvel}^k_\nodept \approx \frac{{F}^k_\nodept-I_3}{h},
        \label{eq:discrete_omega}
    \end{equation}
     which can be interpreted as the first order approximation of $F_\nodept^k = \exp{(\widehat{\angvel_\nodept^k}h)}$.
    After a few matrix manipulations, the discrete rotational kinetic energy in a time interval $[k,k+1]$, spatially discretised in (\ref{eq:rot_ke_sd}), becomes
    \begin{equation}
        \mathcal{T}^k_{rd} = \frac{l_{\nodes}}{h\overline{e}^k_\nodept}\text{tr}[(I_3-{F}^k_\nodept){J}_d],
    \end{equation}
    where $\text{tr}[\cdot]$ is the trace operator, ${J}_d \in \R^{3 \times 3}$ is the non-standard inertia matrix given as
    \begin{equation}
        {J}_d = \frac{1}{2} \text{tr}[{J}]- {J}, 
    \end{equation}
     \begin{equation}
        \overline{e}^k_\nodept := \bigl(\frac{1}{e^k_\nodept}+\frac{1}{e^k_{\nodept+1}}  \bigr)^{-1}
    \end{equation}
    $\forall \nodept = 1,...,\nodes-1$,
    \begin{equation}
        \overline{e}^k_0 = {e^k_0}
    \end{equation}
    for the base node $\nodept = 0$ and 
    \begin{equation}
        \overline{e}^k_\nodes = {e^k_{\nodes-1}}
    \end{equation}
    for the tip node $\nodept = \nodes$.
     \begin{remark}
         The time discretisation of strain potential energies in (\ref{eq:Usd}) and (\ref{eq:Ubd}) is performed using the rectangular rule of approximation. It is simply the evaluation of each of the terms at the time instant $k$ and the total strain potential energy is denoted by $\mathcal{U}^k_{d}$. This is because the potential energy terms do not contain time derivatives and hence do not require discrete-time velocities of the rod.
         \label{remark:discrete_time_pe}
         \end{remark}
 \begin{remark}
     It should be noted that the temporal discretisation performed in this section preserves symplecticity and satisfies the discrete Noether's theorem of the discrete-time Lagrangian flow, as discussed in \cite{Demoures2015}. 
 \end{remark}        

\subsection{Discrete equations of motion of the rod}
\label{sec:discrete_eom}
The discrete Lagrangian of the rod can be written by combining the kinetic and potential energy terms obtained after spatial and time discretisation. The discrete action sum hence obtained is given as
  \begin{align}
        \begin{split}
            \mathcal{S}_d &= \sum_{k = 0}^{N-1}  \frac{l_{\nodes}}{4h}\rho \Bar{A} ||\Delta \x^k_0||^{2} + \frac{l_{\nodes}}{2he^k_0}\text{tr}[(I_3-{F}^k_0){J}_d] + \\
            &  \sum_{k = 0}^{N-1}  \frac{l_{\nodes}}{4h}\rho \Bar{A} ||\Delta \x^k_{\nodes}||^{2} + \frac{l_{\nodes}}{2he^k_{\nodes-1}}\text{tr}[(I_3-{F}^k_{\nodes}){J}_d]     +\\    
         &\sum_{k= 0}^{N-1} \sum_{\nodept= 1}^{\nodes-1} \frac{l_{\nodes}}{2h}\rho \Bar{A} ||\Delta \x^k_\nodept||^{2} + \frac{l_{\nodes}}{2h \overline{e}^k_\nodept}\text{tr}[(I_3-{F}^k_\nodept){J}_d]- \sum_{k= 0}^{N} \sum_{\nodept= 1}^{\nodes} h \mathcal{U}^k_{d},
         \label{eq:discrete_action_sum_rod}
        \end{split} 
    \end{align}
where $\mathcal{U}^k_d$ is defined in Remark \ref{remark:discrete_time_pe}.\\
The discrete equations of motion of the rod with cross-sectional deformation can be obtained by computing the variation of the discrete action sum given in (\ref{eq:discrete_action_sum_rod}). The  discrete translational equations of motion of the rod are given as
\begin{equation}
    \boxed{\frac{\rho \Bar{A}}{h^2}\bigl(\x^{k+1}_\nodept-2\x^k_\nodept+\x^{k-1}_\nodept  \bigr) = -\mathcal{V}^k_\nodept + \mathcal{F}^k_\nodept,} 
     \label{eq:lgvi_lin_eom}
\end{equation}
where
\begin{align}
    \begin{split}
        \mathcal{V}^k_0 := \left. \frac{2}{l_{\nodes}} \Bigl(-\frac{1}{2}\rot_0 \newstrecthstiff{\Bar{\stretchstiff}}{0} (\discaxstrain{\x}{\rot}{0}{l_{\nodes}})-\frac{1}{2} \rot_{1} \newstrecthstiff{\Bar{\stretchstiff}}{0} (\discaxstrainmplusbase{\x}{\rot}{0}{l_{\nodes}})     \Bigr) \right\vert_{k} 
    \end{split}
\end{align}
for the base node $\nodept = 0$,
\begin{align}
    \begin{split}
 \mathcal{V}^k_{\nodes} :=&  \frac{2}{l_{\nodes}} \Bigl( \frac{1}{2} \rot_{\secondlastnode} \newstrecthstiff{\Bar{\stretchstiff}}{\secondlastnode} (\discaxstrain{\x}{\rot}{\secondlastnode}{l_{\nodes}}) \\
 &+ \left. \frac{1}{2} \rot_{\nodes} \newstrecthstiff{\Bar{\stretchstiff}}{\secondlastnode} (\discaxstrainmplus{\x}{\rot}{\nodes}{l_{\nodes}})   \Bigr) \right\vert_{k}
      \end{split}
\end{align}
for the tip node $\nodept = \nodes$,
 \begin{align}
     \begin{split}
          \mathcal{V}^k_\nodept & :=  \frac{1}{l_{\nodes}} \Bigl(-\frac{1}{2}\rot_\nodept \newstrecthstiff{\Bar{\stretchstiff}}{\nodept} (\discaxstrain{\x}{\rot}{\nodept}{l_{\nodes}}) 
          +\frac{1}{2} \rot_{\nodept-1} \newstrecthstiffB{\Bar{\stretchstiff}}{\nodept} (\discaxstrainA{\x}{\rot}{\nodept}{l_{\nodes}})\\
         & \left. - \frac{1}{2} \rot_{\nodept+1} \newstrecthstiff{\Bar{\stretchstiff}}{\nodept} (\discaxstrainmplusint{\x}{\rot}{\nodept}{l_{\nodes}})+
         \frac{1}{2} \rot_\nodept \newstrecthstiffB{\Bar{\stretchstiff}}{\nodept} (\discaxstrainB{\x}{\rot}{\nodept}{l_{\nodes}}) \Bigr) \right\vert_{k}
     \end{split}
 \end{align}
 for interior nodes $\nodept = \{1,2,..., \nodes-1 \}$ and $\mathcal{F}^k_\nodept$ is the force acting on node $\nodept$ at time instant $k$. It can be observed that the left-hand side of (\ref{eq:lgvi_lin_eom}) is the central finite difference approximation of $\Ddot{\x}$ and the term $-\mathcal{V}^k_\nodept$ on the right-hand side is the discretisation of the spatial derivative of potential energy due to stretching and shearing strains given in (\ref{eq:lin_eom}).\\
 The discrete rotational equations of motion of the rod are given as
 \begin{equation}
    \boxed{ 
    \begin{aligned}
       \frac{l_{\nodes}}{{2}{h^2}\overline{e}^k_{\nodept}}(\implicitA{F}{{J}}{\nodept}{k})^{\vee} &- \frac{l_{\nodes}}{{2}{h^2}\overline{e}^k_{\nodept}}(\implicitB{F}{{J}}{\nodept}{k})^{\vee} 
       \\
       &= \mathcal{W}^k_{\nodept} + \mathcal{\tau}^k_{\nodept}, 
    \end{aligned}}
\label{eq:lgvi_rot_eom}
 \end{equation}
where
\begin{align}
 \begin{split}
     \mathcal{W}^k_0 :=& \frac{1}{2}\newstrecthstiff{\Bar{\stretchstiff}}{0}(\discaxstrain{\x}{\rot}{0}{l_{\nodes}}) \times \rot^T_0 \Delta \x_{0} \\
     & \left. + \frac{1}{l_{\nodes}e^3_0}\bigl( (\termAbase{\rot}{0}{k} \widehat{\Bar{\bendstiff} \psi_0} (\widehat{\psi_0}-2I_3))^{(A)}\bigr)^{\vee} \right\vert_{k}
 \end{split}
 \end{align}
 for the base node $\nodept = 0$,
  \begin{align}
     \begin{split}
       \mathcal{W}^k_{\nodes}& := \frac{1}{2}\newstrecthstifftip{\Bar{\stretchstiff}}{\nodes} (\discaxstrainmplus{\x}{\rot}{\nodes}{l_{\nodes}}) \times \rot^T_{\nodes} \Delta \x_{\nodes-1} \\
       & \left. + \frac{1}{l_{\nodes}e^3_{\nodes-1}} \bigl( (\termBtip{\rot}{\nodes}{k} \widehat{\Bar{\bendstiff} {\psi}_{\nodes-1}}(2I_3-\widehat{{\psi}_{\nodes-1}}){\rot^{T}_{\nodes-1}\rot_{\nodes}})^{(A)} \bigr)^{\vee} \right\vert_{k}
       \end{split}
 \end{align}
 for the tip node $\nodept = \nodes$
 \begin{align}
     \begin{split}
         \mathcal{W}^k_{\nodept} &:=\frac{1}{l_{\nodes}} \Bigg\{\frac{1}{2}\newstrecthstiffB{\Bar{\stretchstiff}}{\nodept} (\discaxstrainB{\x}{\rot}{\nodept}{l_{\nodes}}) \times \rot^T_{\nodept} \Delta \x_{\nodept-1} \\
     &+\frac{1}{2}\newstrecthstiff{\Bar{\stretchstiff}}{\nodept}(\discaxstrain{\x}{\rot}{\nodept}{l_{\nodes}}) \times \rot^T_{\nodept} \Delta \x_{\nodept} \\
     &+\frac{1}{l_{\nodes}e^2_{\nodept}}\bigl( (\termA{\rot}{\nodept}{k} \widehat{\Bar{\bendstiff} \psi_{\nodept}} (\widehat{\psi_{\nodept}}-2I_3))^{(A)}\bigr)^{\vee} \\
     &\left. +\frac{1}{l_{\nodes}e^2_{\nodept-1}} \bigl( (\termB{\rot}{\nodept}{k} \widehat{\Bar{\bendstiff} {\psi}_{\nodept-1}}(2I_3-\widehat{{\psi}_{\nodept-1}}){(\rot_{\nodept-1})^T\rot_{\nodept}})^{(A)} \bigr)^{\vee}\Bigg\} \right\vert_{k}
     \end{split}
     \label{eq:W_m_k}
\end{align}
for interior nodes $\nodept = \{1,2,..., \nodes-1 \}$ and $\tau_{\nodept}^k$ is the external moment at node $\nodept$ at time instant $k$. 
\begin{remark}
    Note that the discrete rotational equation of motion given in (\ref{eq:lgvi_rot_eom}) differs from the one given in \cite{Demoures2015} due to the presence of the discrete dilatation terms $\overline{e}^k_{\nodept},e^k_{\nodept}$ and $e^k_{\nodept-1}$. The presence of these additional terms will affect the solution of the implicit rotational update equation 
\begin{equation}
    F^k_{\nodept}{\Bar{J}_d} - {\Bar{J}_d} (F^k_{\nodept})^T = \widehat{{g}^k_\nodept},
    \label{eq:implicit_rot_eq_67}
\end{equation}
where $g^k_{\nodept} \in \R^3$ consists of the rest of the terms given in (\ref{eq:lgvi_rot_eom}).\\ 
The solution of (\ref{eq:implicit_rot_eq_67}) is obtained using the Cayley map in (\ref{eq:cayley}) and the Newton-Raphson method explained in \cite{Lee_2008}.\\
\end{remark}
\section{Numerical results}
\label{sec:numerical_results}
In this section, we present three numerical experiments to validate the LGVI model of the three-dimensional Cosserat rod which includes cross-sectional deformation given by (\ref{eq:lgvi_lin_eom}) and (\ref{eq:lgvi_rot_eom}). The three numerical tests are:
\begin{enumerate}
    \item The flying beam problem, which is a benchmark experiment and will be used to assess the energy-conserving behaviour of the rod as well as the effect of the dilatation factor on its dynamics.
    \item  Pure stretching of a rod, which will be primarily used to investigate the impact of the dilatation factor on the length of the rod and the volume conservation of its spatial elements.
    \item  The general case of bending and stretching of the rod to examine the effects of incorporating the cross-sectional deformation on the factors described in the previous cases.
\end{enumerate}
In all three cases, we will compare the numerical solutions of the standard and the modified models obtained using the Lie group variation integrator technique. The comparison will be made based on the following errors:
\begin{enumerate}
    \item Maximum energy error: 
    The discrete energy of the rod is obtained by using the trapezoidal rule for approximating kinetic and linear strain energies and the rectangular rule for approximating the rotational strain energy. The discrete energy of the modified rod model, described by (\ref{eq:lgvi_lin_eom}) and (\ref{eq:lgvi_rot_eom}), is given as
    \begin{equation}
        \begin{aligned}
    H_m^k &= \sum_{\nodept = 0,\nodes} \frac{\rho \Bar{A} \l_{\nodes}}{4h^2} ||x^{k+1}_\nodept - x^{k}_\nodept||^2 + \sum_{\nodept = 1}^{\nodes-1} \frac{\rho \Bar{A} \l_{\nodes}}{2h^2} ||x^{k+1}_\nodept - x^{k}_\nodept||^2 \\
    & + \sum_{\nodept = 0}^{\nodes} \frac{l_\nodes}{4 h^2} \text{tr}\bigl[(F_\nodept^k-I_3)^T \frac{\Bar{J}_d}{\overline{e}^k_{\nodept}} (F_\nodept^k-I_3)   \bigr] \\
    & + \sum_{\nodept = 0}^{\nodes-1} \Biggl[ \frac{l_\nodes}{4} \Biggl(\Bigl((\rot^k_\nodept)^T \frac{\Delta \x_\nodept^k}{l_\nodes} - e_3\Bigr)^T \frac{\Bar{\stretchstiff}}{e_\nodept^k}\Bigl((\rot^k_\nodept)^T \frac{\Delta \x_\nodept^k}{l_\nodes} - e_3\Bigr) \\
    & + \Bigl((\rot^k_{\nodept+1})^T \frac{\Delta \x_\nodept^k}{l_\nodes} - e_3\Bigr)^T \frac{\Bar{\stretchstiff}}{e_\nodept^k}\Bigl((\rot^k_{\nodept+1})^T \frac{\Delta \x_\nodept^k}{l_\nodes} - e_3\Bigr)\Biggr) \\
    &+\frac{1}{2 l_{\nodes}} (\curvature_\nodept^k)^T \frac{\Bar{\bendstiff}}{(e_\nodept^k)^3} \curvature_\nodept^k  \Biggr]
\end{aligned}
\label{eq:H_m}
\end{equation}
for $k = \{0,...,N\}$. The discrete energy in (\ref{eq:H_m}) will be used to demonstrate the energy-conserving behaviour of the discrete equations of motion given in (\ref{eq:lgvi_lin_eom}) and (\ref{eq:lgvi_rot_eom}). The discrete energy of the rod without the cross-sectional deformation, given in \cite{Demoures2015}, will be denoted by $H_s^k$, and is given by assuming $e_\nodept^k, \Bar{e}_q^k = 1$ for all $\nodept = \{0,...,\nodes$\} and $k = \{0,...,N\}$. \\
    The maximum error in energy between the standard and the modified model is defined as
    \begin{equation}
        \Delta H_{max} = \max_{0 \leq k \leq N} | H_m^k-H_s^k|.
        \label{eq:energy_error}
    \end{equation}
    \item Maximum length error: The maximum error in length between the models is given as
    \begin{equation}
        \Delta L_{max} = \max_{0 \leq k \leq N} \Bigg|\frac{L_m^k-L_s^k}{L_0} \Bigg|
        \label{eq:length_error}
    \end{equation}
    where $L_0$ is the initial length of the rod, $L_s^k$ and $L_m^k$ are the lengths of the standard and modified models of the rod at time instant $k$, respectively.
    \item Maximum volume error: The maximum error in volume between the two models is defined as
    \begin{align}
    \begin{split}
        \Delta V_{max} & = \max_{0 \leq k \leq N, 0 \leq \nodept \leq  \nodes} \Bigg|\frac{(V_\nodept^k)_m-(V_\nodept^k)_s}{V_0} \Bigg| \\
        &= \max_{0 \leq k \leq N, 0 \leq \nodept \leq  \nodes} \Bigg|\frac{{A_\nodept^k}(\Delta \x_\nodept^k)_m-\Bar{A}_\nodept^k (\Delta \x_\nodept^k)_s}{\Bar{A}_\nodept^k l_\nodes} \Bigg|
    \end{split}
    \label{eq:vol_error}
    \end{align}
    where $V_0$ is the initial volume of each element of length $l_\nodes$ of the rod, $(V_\nodept^k)_s $ and $(V_\nodept^k)_m$ are the volumes of the infinitesimal elements $(\Delta \x_\nodept^k)_s$ and $(\Delta \x_\nodept^k)_m$ of the rod described by the standard and the modified models at time instant $k$, respectively.\\
    On substituting the expressions for the cross-sectional deformation and the discrete dilatation factor given in (\ref{eq:modified_area}) and (\ref{eq:discrete_dilatation}), respectively, in (\ref{eq:vol_error}), we get
    \begin{equation}
         \Delta V_{max} = \max_{0 \leq k \leq N, 0 \leq \nodept \leq  \nodes} \Bigg|\frac{l_{\nodes}-(\Delta \x_\nodept^k)_s}{l_{\nodes}} \Bigg| .
        \label{eq:volume_error}
    \end{equation}

\end{enumerate}
\subsection{Flying beam}
\label{sec:flying_beam}
\begin{figure}[H]
    \centering

    \begin{subfigure}{0.50\textwidth}
        \includegraphics[width=\linewidth]{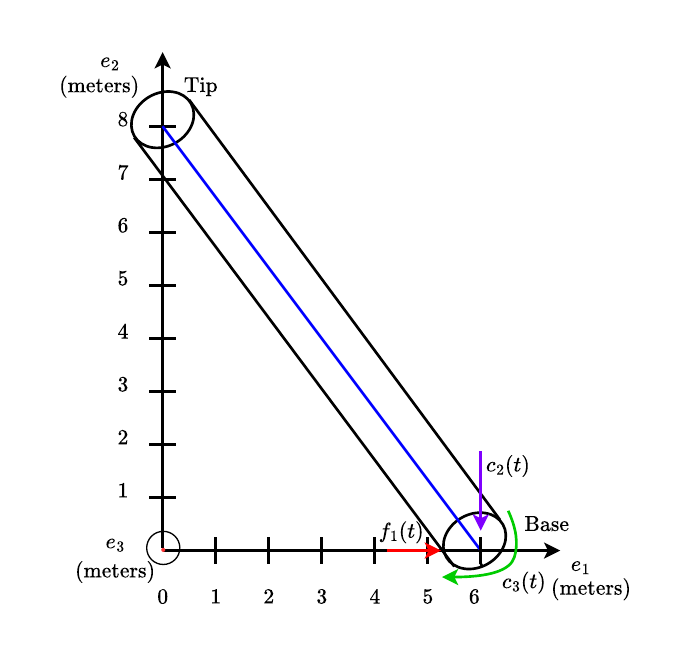}
        \subcaption{}
    \end{subfigure}
    \hfill
    \begin{subfigure}{0.40\textwidth}
        \includegraphics[width=5cm,height=5cm,keepaspectratio]{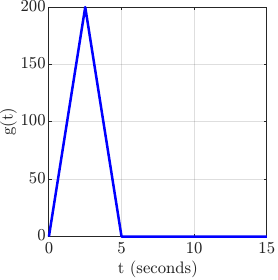}
        \subcaption{}
    \end{subfigure}

    \caption{Flying beam: (a) Initial configuration of the beam. (b) The time-dependent function $g(t)$ is used to define external force and moments at the base of the beam.}
    \label{fig:Hante_config}
\end{figure}

Let us consider the benchmark example of a flying beam in space without the effect of gravity. We consider the parameters of the rod as given in \cite{Hante}. Thus, the flying beam problem consists of a rod of length $10 \hspace{0.2em} \text{meters}$ with the following material properties:
$GA = EA = 10^4 \hspace{0.2em} \text{N}$ , $EI = GJ = 500 \hspace{0.2em} \text{N}\text{m}^2$, $\rho A = 1 \hspace{0.2em} \text{kg}/\text{m}$ and $\rho I = 10 \hspace{0.2em} \text{kg} \hspace{0.1em} \text{m}$. The rod is free at both ends and lies initially inclined in $e_1 - e_2$ plane as shown in Figure \ref{fig:Hante_config}(a). An external force $f_1 (t)$ along the $e_1$ axis and external moments $c_2(t)$ and $c_3(t)$ about the $e_2$ and $e_3$ axes, respectively, are applied at the base of the rod. The external force and moments are described by a time-dependent function $g(t)$ as shown in Figure \ref{fig:Hante_config}(b). The external force and moments are defined as
\begin{align}
    \begin{split}
    f_1 (t) = \frac{1}{10} g(t) {e_1} \hspace{0.2em} \text{N}, \quad c_2 (t) =  -\frac{1}{2} g(t) {e_2} \hspace{0.2em} \text{Nm}, \quad 
    c_3 (t) =  -g(t) {e_3} \hspace{0.2em} \text{Nm} .
    \end{split}
\end{align}
The rod is spatially discretised into $100$ elements and simulated for $15 \hspace{0.2em} \text{seconds}$  with a time step of $10^{-4} \hspace{0.2em} \text{seconds}$ . 
\begin{figure}[H]
    \centering
    \includegraphics[width=\linewidth]{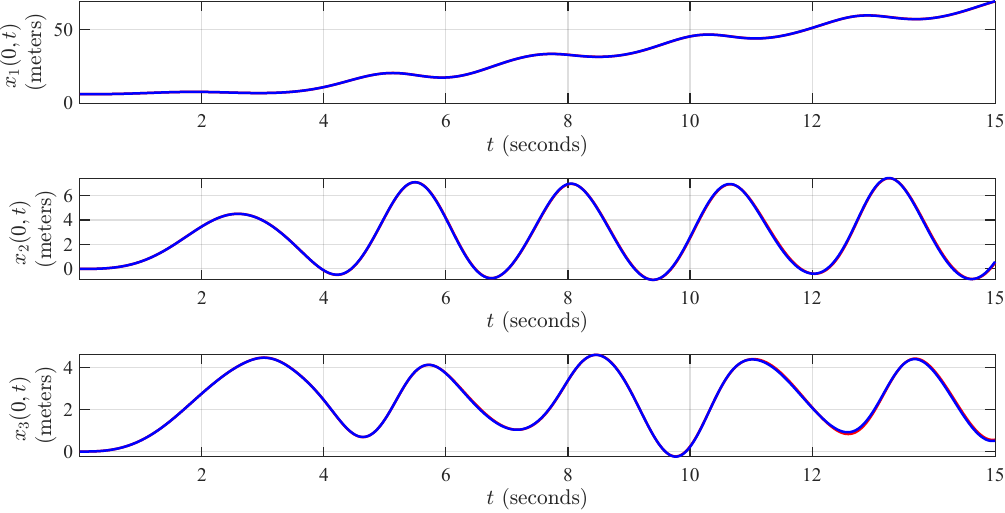}
    \caption{Flying beam: Deflection of the base of the beam for the standard (blue) and the modified (red) models. The component of the position of the 3D beam along $e_i$ is denoted by $x_i$ for $i = 1,2,3$.}
    \label{fig:Hante_deflection}
\end{figure}
\begin{figure}[H]
    \centering
    \includegraphics[width=0.8\linewidth]{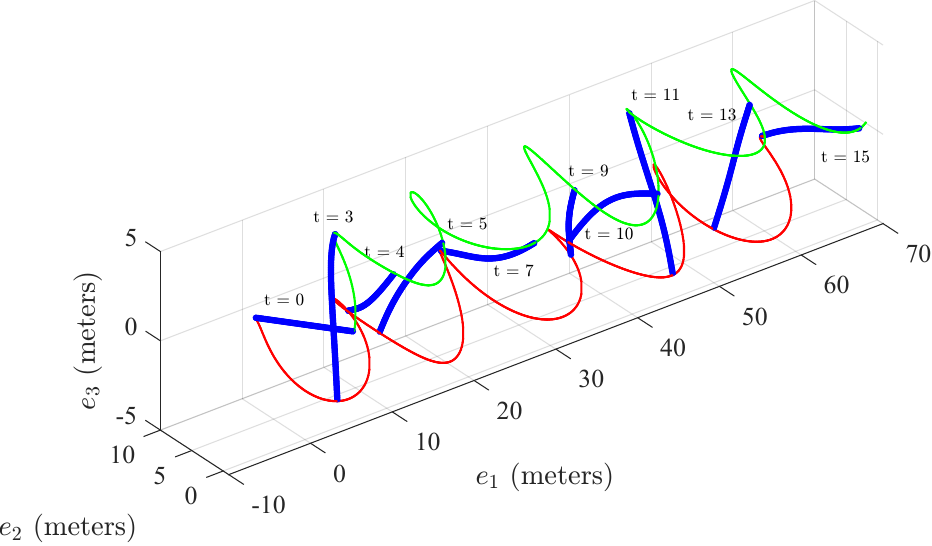}
    \caption{Flying beam: Time evolution of the 3D beam including cross-sectional deformation, base trajectory (green) and tip trajectory (red). }
    \label{fig:Hante_3D_plot}
\end{figure}
    
\begin{figure}[H]
    \centering
     \begin{subfigure}{0.45\textwidth}
    \includegraphics[width=6.5cm,height=4cm,keepaspectratio]{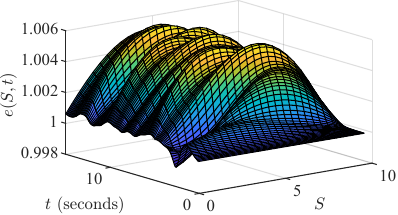}
    \subcaption{}
    \end{subfigure}
    \hfill
    \begin{subfigure}{0.50\textwidth}
    \includegraphics[width=6cm,height=4cm,keepaspectratio]{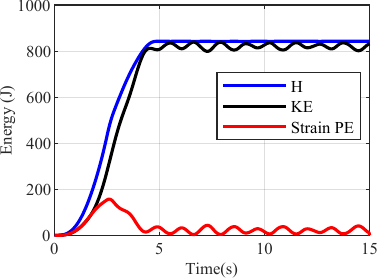}
    \subcaption{}
    \end{subfigure}

    \caption{Flying beam: (a) Evolution of the local dilatation factor $e(S,t)$. (b) Kinetic energy, strain potential energy and total energy of the beam. The energy of the LGVI model is conserved after the removal of external force and moments.  }
    \label{fig:case1_dilataion_energy}
\end{figure}

The deflections obtained using the modified and standard models have an error of 0.0053 meters, thus validating the overlapping curves for the time evolution of the base of the rod shown in Figure \ref{fig:Hante_deflection}. Figure \ref{fig:Hante_3D_plot} demonstrates the time evolution of the rod including the effects of the dilatation factor. The figure also shows the trajectory of the base and the tip nodes.\\
As shown in Figure \ref{fig:case1_dilataion_energy}(a), the local dilatation factor varies for each spatial node and lies in the range $0.9993-1.0059$. Thus, the rod does not experience significant elongation due to the external force and moments. Figure \ref{fig:case1_dilataion_energy}(b) demonstrates the energy behaviour of the flying beam. Since the external force and moments are time-dependent, therefore, the energy is not conserved till $5 \hspace{0.2em} \text{seconds}$. It can be observed from Figure \ref{fig:case1_dilataion_energy}(b) that the modified model exhibits good energy-conserving behaviour after the external force and moments are removed.
The addition of the dilatation factor introduces an error in energy of $\Delta H_{max} = 1.955 \hspace{0.2em} \text{J}$. Thus, the Lie group variational integrator incorporating the cross-sectional deformation conserves the approximate energy of the rod. 
\subsection{Pure stretching}
\label{sec:pure_strectching}
\begin{figure}[H]
    \centering
     \begin{subfigure}{0.45\textwidth}
    \includegraphics[width=\linewidth]{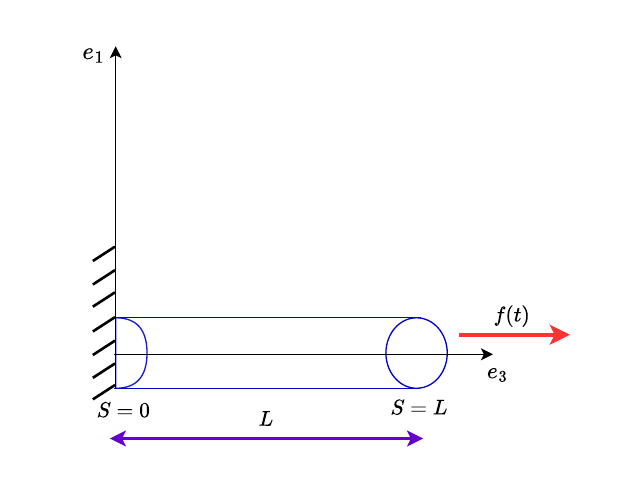}
    \subcaption{}
    \end{subfigure}
    \hfill
    \begin{subfigure}{0.50\textwidth}
    \includegraphics[width=10cm,height=4cm,keepaspectratio]{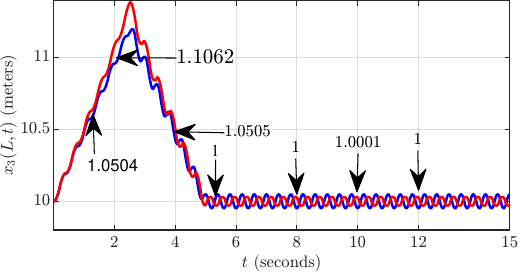}
    \subcaption{}
    \end{subfigure}
    
   \caption{Pure stretching: (a) Initial condition of the rod and the axial force. (b) Tip deflection of the standard (blue) and the modified (red) models and values of the dilatation factor at various time instants.}
    \label{fig:case2_config_deflection}
\end{figure}

\begin{figure}[H]
    \centering
     \begin{subfigure}{0.45\textwidth}
    \includegraphics[width=6cm,height=6cm,keepaspectratio]{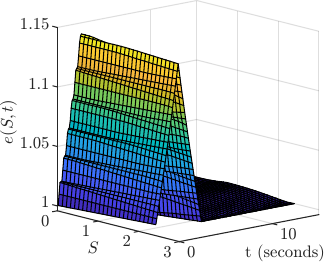}
    \subcaption{}
    \end{subfigure}
    \hfill
    \begin{subfigure}{0.45\textwidth}
    \includegraphics[width=6cm,height=6cm,keepaspectratio]{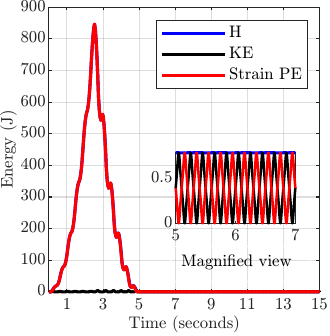}
    \subcaption{}
    \end{subfigure}
   \caption{Pure stretching: (a) Evolution of the dilatation factor at various nodes. (b) Kinetic energy, strain potential energy and total energy of the rod incorporating cross-sectional deformation. The total energy is conserved after the axial force is removed at $5 \hspace{0.2em} \text{seconds}$ as shown in the magnified view. }
    \label{fig:case2_dilatation_energy}
\end{figure}

Consider a cantilever rod undergoing pure stretching to analyse the effects of the local dilatation factor on the dynamics of the rod. We consider the rod of length $L = 10 \hspace{0.2em} \text{meters}$ with the geometrical and material-specific properties as described in the previous case. The rod initially lies along the $e_3$ axis as shown in Figure \ref{fig:case2_config_deflection}(a). An axial force of $6 g(t) \hspace{0.2em} \text{N}$ is applied at the tip of the rod, where $g(t)$ is the time-dependent function as shown in Figure \ref{fig:Hante_config}(b). The rod is discretised into 100 spatial elements and is simulated for $15 \hspace{0.2em} \text{seconds}$ with a time step of $10^{-4} \hspace{0.2em} \text{seconds}$. Figure \ref{fig:case2_config_deflection}(b) shows the deflection of the tip of the rod. It can be observed that the rod undergoes stretching for $t \in [0,5] \hspace{0.2em} \text{seconds}$. The rod oscillates about it's initial configuration after the removal of the external axial force. The time evolution of the local dilatation factor for all spatial nodes is shown in Figure \ref{fig:case2_dilatation_energy}(a). The plot indicates that $e(S,t)$ varies in the range $0.9920-1.1408$. This additional factor significantly affects of the length of the rod and the volume of its discrete elements as compared to the standard model, with the errors $\Delta L_{max} = 0.02070$ and $\Delta V_{max} = 0.1247$. Figure \ref{fig:case2_dilatation_energy}(b) depicts the energy behaviour of the modifed model of the rod. The error in total energy between the two models is $148.77 \hspace{0.2em} \text{J}$. It can be observed from the figure that the approximate energy of the rod is conserved after the removal of the external force. Thus, the discrete modified model captures the planar cross-sectional deformation as well as the energy-conserving behaviour of the Cosserat rod.  

\subsection{Bending and stretching}
\label{sec:bending_and_strecthing}
\begin{figure}[H]
    \centering
     \begin{subfigure}{0.45\textwidth}
    \includegraphics[width=\linewidth]{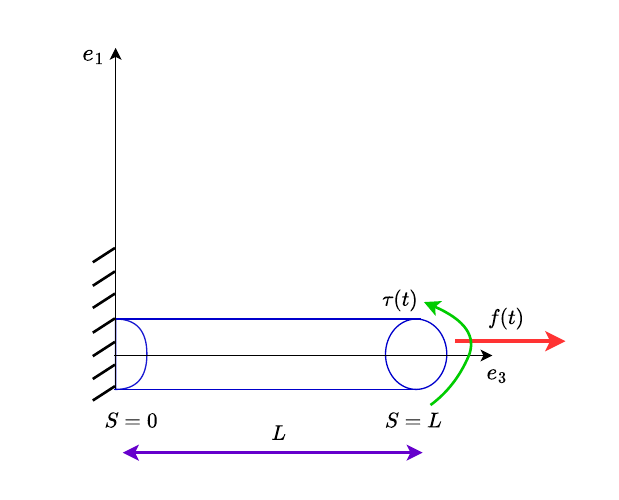}
    \subcaption{}
    \end{subfigure}
    \hfill
    \begin{subfigure}{0.50\textwidth}
    \includegraphics[width=7cm,height=5cm,keepaspectratio]{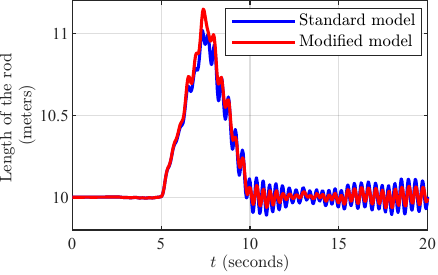}
    \subcaption{}
    \end{subfigure}
   \caption{Bending and stretching: (a) Initial condition of the rod, the axial force and the bending moment applied at the tip. (b) Length of the rod for the two models at various time instants.}
    \label{fig:case3_config_length}
\end{figure}

\begin{figure}[H]
    \centering
     \begin{subfigure}{0.45\textwidth}
    \includegraphics[width=6cm,height=6cm,keepaspectratio]{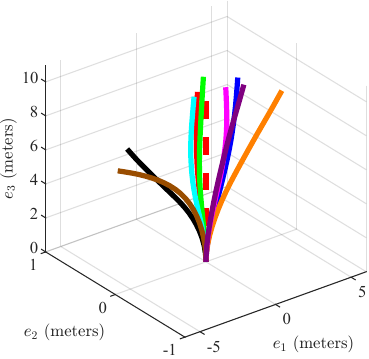}
    \subcaption{}
    \end{subfigure}
    \hfill
    \begin{subfigure}{0.40\textwidth}
    \includegraphics[width=6cm,height=6cm,keepaspectratio]{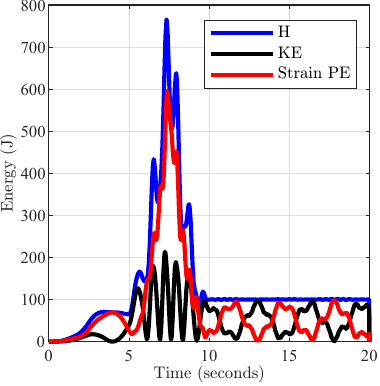}
    \subcaption{}
    \end{subfigure}
   \caption{Bending and stretching: (a) Deflection of the modified model of the cantiliver rod at the following time instants: $t = 0 \hspace{0.2em} \text{seconds}$  (red, dotted line), $t = 5 \hspace{0.2em} \text{seconds}$  (red), $t = 6 \hspace{0.2em} \text{seconds}$ (dark blue), $t = 8 \hspace{0.2em} \text{seconds}$ (green), $t = 10 \hspace{0.2em} \text{seconds}$ (light blue), $t = 12 \hspace{0.2em} \text{seconds}$ (black), $t = 14 \hspace{0.2em} \text{seconds}$ (orange), $t = 16 \hspace{0.2em} \text{seconds}$ (magenta), $t = 18 \hspace{0.2em} \text{seconds}$ (brown) and $t = 20 \hspace{0.2em} \text{seconds}$ (violet). (b) Time evolution of kinetic energy, strain potential enery and total energy of the rod including cross-sectional deformation. The total energy is conserved with a bounded error after the removal of the external force and moment from $10 \hspace{0.2em} \text{seconds}$ to $20 \hspace{0.2em} \text{seconds}$ .}
    \label{fig:case3_delfection_energy}
\end{figure}

Consider a cantilever rod of length $10 \hspace{0.2em} \text{meters}$ initially lying along the $e_3$ axis as shown in Figure \ref{fig:case3_config_length}(a). The geometrical and material-specific properties are given in Section \ref{sec:flying_beam}. The rod is subjected to an external force $f(t)$ and an external moment $\tau (t)$ at the tip. The numerical test considered in this section illustrates the effects of bending moment and axial force on the dynamics of the rod. This is particularly useful in modelling of soft robots, for example, an octopus-inspired soft robotic arm performing reaching and tracking tasks. The external force is given by $f(t) = 5q(t) e_3 \hspace{0.2em} \text{N}$, where $q(t)$ is defined as
\begin{equation}
    q(t) = \begin{cases}
        0, & 0 \leq t \leq 5\\
        80(t-5), & 5 < t \leq 7.5 \\
        80(10-t), & 7.5< t \leq 10\\
        0, & t > 10
    \end{cases}.
\end{equation}
The tip moment is given as $\tau(t) = -\frac{1}{2} g(t) e_2 \hspace{0.2em} \text{Nm} $, where $g(t)$ is described in Figure \ref{fig:Hante_config}(b). The rod is spatially discretised coarsely into 20 elements and is simulated for $20 \hspace{0.2em} \text{seconds}$ with a time step of $10^{-4} \hspace{0.2em} \text{seconds}$. Figure \ref{fig:case3_config_length}(b) shows the length of the standard and modified models of the rod with the error $\Delta L_{max} = 0.0182$. The three-dimensional shape of the rod at various time instants is shown in Figure \ref{fig:case3_delfection_energy}(a). It can be observed that the rod obtains interesting shapes due to the application of external force and moment. The local dilatation factor of the rod which varies in the range $0.9688-1.1207$, ensures volume conservation of the discrete elements of the rod incorporating cross-sectional deformation with $\Delta V_{max} = 0.1073$. Figure \ref{fig:case3_delfection_energy}(b) shows the energy behaviour of the rod. The sharp increase in the energy for $t \in [5,10] \hspace{0.2em} \text{seconds}$ is due to the axial force applied at the tip of the rod. The energy is almost conserved after the removal of external force and moment for $t \in [10,20] \hspace{0.2em} \text{seconds}$. Therefore, the LGVI model ensures volume conservation of the discrete elements of the rod and validates energy conservation with a bounded error.
\subsection{Numerical convergence analysis}
\label{sec:convergence_analysis}
In this section, we discuss the numerical convergence of the Lie group variational integrator described by (\ref{eq:lgvi_lin_eom}) and (\ref{eq:lgvi_rot_eom}). We consider a cantilever rod of length $10 \hspace{0.2em} \text{meters}$ initially lying along the $e_3$ axis as shown in Figure \ref{fig:case3_config_length}(a). The geometrical and material-specific properties are given in Section \ref{sec:flying_beam}. The rod is subjected to a simultaneous external force $f(t)$ and an external moment $\tau(t)$ at the tip. The external force and moment are given by $f(t) = 5u(t) {e_3}\hspace{0.2em} \text{N}$ and $\tau(t) = \frac{-1}{2}u(t){e_2} \hspace{0.2em} \text{Nm}$, respectively, where $u(t)$ is defined as 
\begin{equation}
    q(t) = \begin{cases}
        200t, & 0 \leq t \leq 1\\
        -200(t-2), & 1 < t \leq 2 \\
        0, & t>2
    \end{cases}.
\end{equation}
The convergence in space and time is analysed for a simulation time of 2 seconds by means of the errors in position and orientation defined as
\begin{equation}
    \epsilon_{\text{pos}} = \max_{0 \leq k \leq N, 0 \leq \nodept \leq \nodes} \frac{||\x_\nodept^k - (\x_\nodept^k)_{\text{ref}}||_2}{||(\x_\nodept^k)_{\text{ref}}||_2},
\end{equation}
\begin{equation}
    \epsilon_{\text{rot}} = \max_{0 \leq k \leq N, 0 \leq \nodept \leq \nodes} ||I_3 - (\rot_\nodept^k)^T_{\text{ref}}\rot_\nodept^k||_{F},
\end{equation}
where $||.||_2$ is the Euclidean norm, $||.||_F$ is the Frobenius norm, $(\x_\nodept^k)_{\text{ref}} \in \R^3$ and $(\rot_\nodept^k)_{\text{ref}} \in SO(3)$ are the position and orientation for the reference configuration (as described in points \ref{space_convergence} and \ref{time_convergence} given below) and $(\x_\nodept^k) \in \R^3$ and $(\rot_\nodept^k) \in SO(3)$ are the position and orientation for the current configuration. 
\begin{enumerate}
    \item \label{space_convergence}Convergence in space: We consider a fixed time step of $h = 10^{-4}$ seconds and various spatial discretisations with number of elements $\nodes = \{10,100,200,300,400,500,600,700,800\}$. The reference configuration is considered with 900 spatial elements. The errors in position and orientation are shown on log plots in Figure \ref{fig:space_con} (a) and Figure \ref{fig:space_con} (b), demonstrating the decrease in the errors as the spatial grid is refined.
    \begin{figure}[H]
    \centering
     \begin{subfigure}{0.45\textwidth}
    \includegraphics[width=6cm,height=7cm,keepaspectratio]{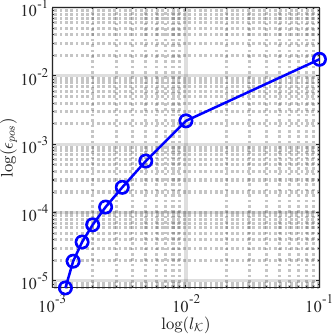}
    \subcaption{}
    \end{subfigure}
    \hfill
    \begin{subfigure}{0.50\textwidth}
    \includegraphics[width=6cm,height=7cm,keepaspectratio]{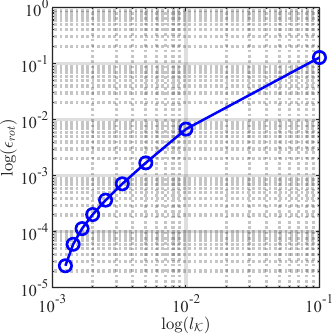}
    \subcaption{}
    \end{subfigure}
   \caption{Convergence in space with fixed time step $h = 10^{-4}$: (a) $\epsilon_{\text{pos}}$ vs $\l_\nodes$. (b) $\epsilon_{\text{rot}}$ vs $\l_\nodes$ for $\nodes = \{10,100,200,300,400,500,600,700,800\}$. Base-10 logarithm has been considered.}
    \label{fig:space_con}
\end{figure}
    \item \label{time_convergence}Convergence in time: We consider 50 spatial elements and various time steps $h = \{10^{-5},5 \times 10^{-5},10^{-4}, 5 \times 10^{-4}, 10^{-3} \}$ seconds. The reference configuration is considered at $h = 10^{-6}$ seconds. The errors in position and orientation are shown on log plots in Figure \ref{fig:time_con} (a) and Figure \ref{fig:time_con} (b), demonstrating the decrease in the errors as the temporal grid is refined.
    \begin{figure}[H]
    \centering
     \begin{subfigure}{0.45\textwidth}
    \includegraphics[width=6cm,height=7cm,keepaspectratio]{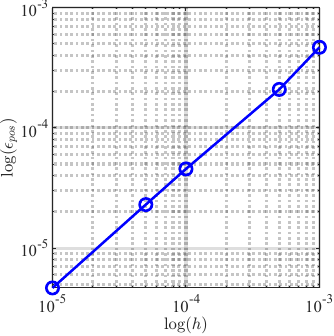}
    \subcaption{}
    \end{subfigure}
    \hfill
    \begin{subfigure}{0.50\textwidth}
    \includegraphics[width=6cm,height=7cm,keepaspectratio]{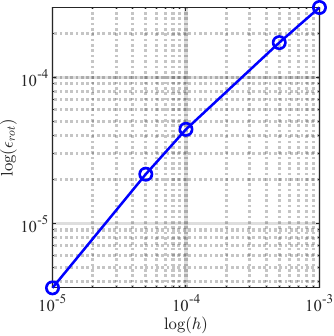}
    \subcaption{}
    \end{subfigure}
   \caption{Convergence in time with a fixed number spatial elements $\nodes = 50$: (a) $\epsilon_{\text{pos}}$ vs $h$. (b) $\epsilon_{\text{rot}}$ vs $h$ for $h = \{10^{-5},5 \times 10^{-5},10^{-4}, 5 \times 10^{-4}, 10^{-3} \}$ seconds. Base-10 logarithm has been considered.}
    \label{fig:time_con}
\end{figure}
\end{enumerate}
\section{Conclusions and Future Work}
\label{sec:conclusion_and_future_work}
In this paper, we derived the continuous space-time equations of motion of the three-dimensional Cosserat rod which incorporated planar cross-sectional deformation via the local dilatation factor $e(S,t)$. Furthermore, the discrete model of the Cosserat rod was developed using the Lie group variational integrator technique and three numerical tests were presented which demonstrated the properties of the proposed discrete model: volume conservation of the discrete elements of the rod and energy conservation with a bounded error. The convergence analysis in space and time demonstrated numerical stability of the proposed integrator and energy conservation for coarser grids in space and time. Future work includes the design of model-based control using the proposed discrete equations of motion for soft robots modelled using the Cosserat rod.

\appendix
\section{Continuous space-time equations of motion of the Cosserat rod incorporating cross-sectional deformation}
\label{app:derivation_CT}
In this section, the translational and rotational dynamics of the Cosserat rod given in (\ref{eq:lin_eom}) - (\ref{eq:rot_bc}), are derived by evaluating the variation of the kinetic and strain potential energies present in the action sum in (\ref{eq:action_int}).\\
The perturbation of the position of the rod, due to the variation $\delta \x (S,t) \in \R^3$ is given as
\begin{equation}
    \x^{\alpha} (S,t) = \x (S,t) + \alpha \delta \x (S,t),
\end{equation}
where $\alpha \in \R$.\\
The perturbation of the orientation of the rod is given as
\begin{equation}
    \rot^{\alpha} (S,t) = \rot (S,t) \exp{(\alpha \widehat{\eta}(S,t))},
\end{equation}
where $\eta (S,t) \in \R^3$. Thus,
\begin{equation}
    \delta \rot (S,t) = \frac{d}{d \alpha}|_{\alpha = 0} \rot^{\alpha} (S,t) = \rot (S,t) \widehat{\eta}(S,t).
\end{equation}
We will now derive the variations of the kinetic and strain potential energies of the rod:
\begin{enumerate}
\item Variation of the translational kinetic energy in (\ref{eq:action_int}): The variation of the translational kinetic energy present in the action integral is given as
\begin{equation}
\begin{aligned}
     \delta \int_{0}^{T} \mathcal{T}_l \,dt &= \delta \int_{0}^{T} \int_{0}^{L} \frac{1}{2} \rho \Bar{A}(S) \Dot{\x}^T(S,t) \Dot{\x} (S,t) \,dS \,dt \\
     & = \int_{0}^{L} \int_{0}^{T} \rho \Bar{A}(S) \Dot{\x}^T(S,t) \delta \Dot{\x} (S,t) \,dt \,dS.  \end{aligned}
\label{eq:variation_lin_KE_1}
\end{equation}
On integrating by parts and substituting $\delta \x(S,0) = \delta \x(S,T) = 0 \in \R^3 \hspace{0.5em} \forall S \in [0,L] $ , we get
\begin{equation}
    \begin{aligned}
       \delta \int_{0}^{T} \mathcal{T}_l \,dt =  -\int_{0}^{T} \int_{0}^{L} \rho \Bar{A}(S)\Ddot{\x}^T (S,t) \delta \x (S,t) \,dS \,dt.
    \end{aligned}
\label{eq:variation_lin_KE}
\end{equation}
\item Variation of the rotational kinetic energy in (\ref{eq:action_int}): The variation of the rotational kinetic energy present in the action integral is given as
\begin{equation}
    \begin{aligned}
        \delta \int_{0}^{T} \mathcal{T}_r (t) \,dt = \delta \int_{0}^{T} \int_{0}^{L} \frac{1}{2} \rho {\angvel^T (S,t)} \frac{\overline{\secondMoI} (S)}{e(S,t)} \angvel(S,t) \, dS  \,dt.
    \end{aligned}
\end{equation}
Considering the variation $\angvel (S,t) + \alpha \delta \angvel (S,t) \in \R^3$ for all $S \in [0,L]$, we get 
\begin{equation}
    \begin{aligned}
        \delta \int_{0}^{T} \mathcal{T}_r (t) \,dt &= \int_{0}^{L} \int_{0}^{T} \rho {\angvel^T (S,t)} \frac{\overline{\secondMoI} (S)}{e(S,t)} \delta \angvel(S,t)  \,dt \,dS.
    \end{aligned}
\end{equation}
On substituting $\delta \angvel (S,t) = \Dot{\eta} (S,t) + \angvel(S,t) \times \eta (S,t)$ and  integrating by
parts, we get
\begin{equation}
\begin{aligned}
    \delta \int_{0}^{T} \mathcal{T}_r (t) \,dt = &\int_{0}^{T} \int_{0}^{L} \bigl( \angvel^T(S,t)\frac{\rho \Bar{\mathcal{I}}(S)}{e(S,t)} \widehat{\angvel}(S,t) - \Dot{\angvel}(S,t)\frac{\rho \Bar{\mathcal{I}}(S)}{e(S,t)} \\ 
    &+ \angvel^T(S,t) \frac{\rho \Bar{\mathcal{I}}(S)}{e^2(S,t)} \frac{de (S,t)}{dt}   \bigr) \eta(S,t) \,dS \,dt.
\end{aligned}
\label{eq:variation_rot_KE}
\end{equation}

Note that while deriving \eqref{eq:variation_rot_KE}, we should in principle consider the variation in the second moment of inertia $\secondMoI$ caused by the variation in $\x$. Considering it would lead to an additional term in the translational equation of motion, which is due to a radial effect in the rod. In the current work, since we consider a slender rod, we have ignored such radial effects, which is consistent with the derivation of the translational equation of motion in \cite{gazzola_2018}.


    \item Variation of linear strain potential energy in (\ref{eq:action_int}):
    The perturbations of linear strains due to $\delta x (S,t)$ and $\delta R (S,t)$ are given as
    \begin{equation}
      \begin{aligned}
        & \epsilon_{a}^{\alpha} = \epsilon_{a} + \alpha \delta \epsilon_{a}, \\
        & \epsilon_{s_1}^{\alpha} = \epsilon_{s_1} + \alpha \delta \epsilon_{s_1}, \\
        & \epsilon_{s_2}^{\alpha} = \epsilon_{s_2} + \alpha \delta \epsilon_{s_2}.
      \end{aligned}
    \end{equation}
The perturbation of the linear strain potential energy present in the action integral in (\ref{eq:action_int}) is given as
\begin{equation}
\begin{aligned}
    \mathcal{U}_{lin}^{\alpha} & = \int_{0}^{T} \int_{0}^{L} \bigl[ \int_{0}^{\epsilon_{s_1}+ \alpha \delta \epsilon_{s_1}} f_1 (\xi_{s_1})\,d\xi_{s_1} + \int_{0}^{\epsilon_{s_2} + \alpha \delta \epsilon_{s_2}} f_2 (\xi_{s_2})\,d\xi_{s_2} \\
    & + \int_{0}^{\epsilon_{a}+ \alpha \delta \epsilon_{a}}f_3 (\xi_{a})\,d\xi_{a}  \bigr] \,dS \,dt.
\end{aligned}
\end{equation}
The variation in the linear strains is given as
\begin{equation}
\begin{aligned}
    \delta V(S,t) &= \begin{bmatrix}
        \delta \epsilon_{s_1} \\
        \delta \epsilon_{s_2} \\
        \delta \epsilon_{a}
    \end{bmatrix} = \delta \Bigl(\rot^T(S,t) \frac{\partial \x (S,t)}{\partial S}\Bigr) \\
    &= \rot^T(S,t) \delta \x^{'}(S,t) + V(S,t) \times \eta(S,t).
\end{aligned}
\label{eq:delta_V}
\end{equation}
The variation in the linear strain potential energy is
\begin{equation}
    \delta \mathcal{U}_{lin} = \frac{d}{d \alpha}|_{\alpha = 0} \hspace{0.2em} \mathcal{U}_{lin}^{\alpha}.
\end{equation}
By the fundamental theorem of calculus, we get
\begin{equation}
    \delta \mathcal{U}_{lin} = \int_{0}^{T} \int_{0}^{L}  f_1 (\epsilon_{s_1}) \delta \epsilon_{s_1} + f_2 (\epsilon_{s_2}) \delta \epsilon_{s_2} + f_3 (\epsilon_{a}) \delta \epsilon_{a}  \,dS \,dt.
    \label{eq:delta_W_lin}
\end{equation}
On substituting $f_1(\epsilon_{s_1}),f_2(\epsilon_{s_2}),f_3(\epsilon_{a})$ from (\ref{eq:fs1}) - (\ref{eq:fa}) and $\delta V(S,t)$ from (\ref{eq:delta_V}) in (\ref{eq:delta_W_lin}), we get
\begin{equation}
\begin{aligned}
    \delta \mathcal{U}_{lin} = &\int_{0}^{T} \int_{0}^{L} (V(S,t)-e_3)^T \frac{\stretchstiff}{e(S,t)}  (\rot^T(S,t) \delta \x^{'}(S,t) \\
    &+ \widehat{V}(S,t) \eta(S,t)) \,dS \,dt.
\end{aligned}
\end{equation}
The term in the translational equation due to $\delta \mathcal{U}_{lin}$ is 
\begin{equation}
    \mathcal{I}_1 = \int_{0}^{T} \int_{0}^{L} (V(S,t)-e_3)^T \frac{\Bar{\stretchstiff}(S)}{e(S,t)}  \rot^T(S,t) \delta \x^{'}(S,t) \,dS \,dt,
\end{equation}
where $\Bar{\stretchstiff}(S) = \text{diag}(G \area,G \area,E \area)$
On integrating by parts, we get
\begin{equation}
\begin{aligned}
    \mathcal{I}_1 &= \int_{0}^{T} \left. \Bigl(R(S,t)\frac{\Bar{\stretchstiff}(S)}{e(S,t)}(V(S,t)-e_3) \Bigr)^T \delta \x(S,t)\right\vert_{S = 0}^{L} \,dt \\
    &- \int_{0}^{T} \int_{0}^{L} \partialwrtS \bigl(R(S,t)\frac{\Bar{\stretchstiff}(S)}{e(S,t)}(V(S,t)-e_3)\bigr)^T \delta \x(S,t) \,dS \,dt.
\end{aligned}
\label{eq:lin_PE_act_int1}
\end{equation}
The first and second terms of $\mathcal{I}_1$ contribute to the boundary conditions and translational equations for the rod, respectively.\\
The term in the rotational dynamics due to $\delta \mathcal{U}_{lin}$ is 
\begin{equation}
    \mathcal{I}_2 = \int_{0}^{T} \int_{0}^{L} (V(S,t)-e_3)^T \frac{\stretchstiff}{e(S,t)} \widehat{V}(S,t) \eta(S,t) \,dS \,dt.
    \label{eq:lin_PE_act_int2}
\end{equation}
\item Variation of angular strain potential energy in (\ref{eq:action_int}): The perturbations of angular strains due to $\delta \rot (S,t)$ are given as
\begin{equation}
      \begin{aligned}
        & \curvature_1^{\alpha} = \curvature_1 + \alpha \delta \curvature_1, \\
        & \curvature_2^{\alpha} = \curvature_2 + \alpha \delta \curvature_2, \\
        & \curvature_3^{\alpha} = \curvature_3 + \alpha \delta \curvature_3.
      \end{aligned}
    \end{equation}
The perturbation of the rotational strain potential energy present in the action integral (\ref{eq:action_int}) is 
\begin{equation}
    \begin{aligned}
        \mathcal{U}_{ang}^{\alpha} &= \int_{0}^{T} \int_{0}^{L} \bigl[ \int_{0}^{\curvature_1 + \alpha \delta \curvature_1} f_4 (\xi_{4})\,d\xi_{4} + \int_{0}^{\curvature_2 + \alpha \delta \curvature_2} f_5 (\xi_{5})\,d\xi_{5} \\
        & +\int_{0}^{\curvature_3 + \alpha \delta \curvature_3} f_6 (\xi_{6})\,d\xi_{6}  \bigr] \, dS \,dt.
    \end{aligned}
\end{equation}
The variation of the angular strains is given as
\begin{equation}
 \begin{aligned}
     \delta \curvature (S,t) &= \begin{bmatrix}
        \delta \curvature_1 \\ 
        \delta \curvature_2 \\
        \delta \curvature_3
    \end{bmatrix} = \delta \Bigl(\rot^T(S,t) \frac{\partial \rot (S,t)}{\partial S}   \Bigr) \\
    &  = \eta^{'}(S,t) + \widehat{\curvature}(S,t) \eta(S,t).  
 \end{aligned}
\label{eq:variation_curvature}
\end{equation}
The variation in the rotational strain potential energy is
\begin{equation}
    \delta \mathcal{U}_{ang} = \frac{d}{d \alpha}|_{\alpha = 0} \hspace{0.2em} \mathcal{U}_{ang}^{\alpha}.
\end{equation}
By the fundamental theorem of calculus, we get
\begin{equation}
    \delta \mathcal{U}_{ang} = \int_{0}^{T} \int_{0}^{L}  f_4 (\xi_4) \delta \xi_4 + f_5 (\xi_5) \delta \xi_5 + f_6 (\xi_6) \delta \xi_6  \,dS \,dt.
    \label{eq:delta_U_ang}
\end{equation}
On substituting $ f_4 (\xi_4),f_5 (\xi_5),f_6 (\xi_6)$ from (\ref{eq:fxi4}) - (\ref{eq:fxi6}) and $\delta \curvature(S,t)$ from (\ref{eq:variation_curvature}) in (\ref{eq:delta_U_ang}), and integrating by parts, we get
\begin{equation}
    \begin{aligned}
        \delta \mathcal{U}_{ang}& = \int_{0}^{T} \left. \Bigl( \curvature^T(S,t) \frac{\Bar{\bendstiff}(S)}{e^3(S,t)} \eta (S,t)\Bigr) \right\vert_{0}^{L} \,dt \\
        & -\int_{0}^{T} \int_{0}^{L} \partialwrtS \Bigl(\curvature^T(S,t) \frac{\Bar{\bendstiff}(S)}{e^3(S,t)}  \Bigr)\eta (S,t) \,dS \,dt \\
        & + \int_{0}^{T} \int_{0}^{L} \curvature^T(S,t) \frac{\Bar{\bendstiff}(S)}{e^3(S,t)} \hat{\curvature} (S,t)  \eta (S,t) \,dS \,dt.
    \end{aligned}
    \label{eq:ang_PE_act_int}
\end{equation}
\end{enumerate}
Equations (\ref{eq:variation_lin_KE}), (\ref{eq:lin_PE_act_int1}) contribute to the translational equation of motion given in (\ref{eq:lin_eom}) and (\ref{eq:variation_rot_KE}),(\ref{eq:lin_PE_act_int2}) and (\ref{eq:ang_PE_act_int}) contribute to the rotational equation of motion given in (\ref{eq:rot_eom}).


\bibliographystyle{unsrtnat}
\bibliography{References}




\end{document}